\documentclass[11pt,nofootinbib,showpacs]{revtex4}
\usepackage{amsfonts,amssymb,amsmath,graphicx,multirow,dsfont}
\def\dac{\displaystyle\frac}

\def\[{\left[}
\def\]{\right]}
\def\({\left(}
\def\){\right)}

\def\toto{\leftrightarrow}
\def\ot{\leftarrow}

\newcommand{\diag}{\mathop{\rm diag}\nolimits}

\begin{document}

\baselineskip7mm

\title{Realistic compactification in spatially flat vacuum cosmological models in cubic Lovelock gravity: Low-dimensional case}

\author{Sergey A. Pavluchenko}
\affiliation{Programa de P\'os-Gradua\c{c}\~ao em F\'isica, Universidade Federal do Maranh\~ao (UFMA), 65085-580, S\~ao Lu\'is, Maranh\~ao, Brazil}

\begin{abstract}
In this paper we begin to perform systematical investigation of all possible regimes in spatially flat vacuum cosmological models in cubic Lovelock gravity. We consider the spatial section
to be a product of three- and extra-dimensional isotropic subspaces, with the former considered to be our Universe. As the equations of motion are different for $D=3, 4, 5$ and
general $D \geqslant 6$ cases, we considered them all separately. Due to the quite large amount different subcases, in the current paper we consider only $D=3, 4$ cases.
For each $D$ case we found values for $\alpha$ (Gauss-Bonnet coupling) and $\beta$ (cubic Lovelock coupling)
which separate different dynamical cases, all isotropic and
anisotropic exponential solutions, and study the dynamics in each region to find all possible regimes for all possible initial conditions and any values of $\alpha$ and $\beta$.
The results suggest that in both $D$ cases the regimes with realistic compactification originate from so-called ``generalized Taub'' solution. The endpoint of the
compactification regimes is either anisotropic exponential (for $\alpha > 0$, $\mu \equiv \beta/\alpha^2 < \mu_1$ (including entire $\beta < 0$)) or standard low-energy Kasner regime
(for $\alpha > 0$, $\mu > \mu_1$); as it is compactification regime, both endpoints have expanding three and contracting extra dimensions.
There are two unexpected observations among the results -- all realistic compactification regimes exist only for $\alpha > 0$ and there is no smooth transition between high-energy
and low-energy Kasner regimes, the latter with realistic compactification.
\end{abstract}


%04.20 - General Relativity
%04.20.Dw    Singularities and cosmic censorship
%04.20.Fy    Canonical formalism, Lagrangians, and variational principles
%04.20.Jb    Exact solutions

%04.25.dc    Numerical studies of critical behavior, singularities, and cosmic censorship

%04.50.-h    Higher-dimensional gravity and other theories of gravity
%04.50.Gh    Higher-dimensional black holes, black strings, and related objects
%04.50.Kd    Modified theories of gravity

%11.25Mj     Compactifications and four-dimensional models

%98.80.-k    Cosmology
%98.80.Bp    Origin and formation of the Universe
%98.80.Cq    Particle-theory and field-theory models of the early Universe

\maketitle

\section{Introduction}

It is interesting to note that the idea of extra dimensions is older then General Relativity (GR) itself. Indeed, the first ever extra-dimensional model was constructed by
Nordstr\"om in 1914~\cite{Nord1914}, and it unified Nordstr\"om's second gravity theory~\cite{Nord_2grav} with Maxwell's electromagnetism. After Einstein proposed GR~\cite{einst},
it still took years before it was accepted: during the solar eclipse of 1919, the
bending of light near the Sun was measured and the
deflection angle was in perfect agreement with GR, while Nordstr\"om's theory, as most of the scalar gravity theories, predicted a zeroth deflection angle.

Though, the idea of extra dimensions was not forgotten -- in 1919 Kaluza proposed~\cite{KK1} a similar model but based on GR: in his model five-dimensional Einstein equations could be decomposed into $4D$ Einstein equations and Maxwell's electromagnetism. For such decomposition to exist, the extra dimensions should be ``curled'' or compactified into a circle and ``cylindrical conditions'' should be imposed. The work by Kaluza was followed by Klein who proposed~\cite{KK2, KK3} a nice quantum mechanical interpretation of this extra dimension and so the theory, called Kaluza-Klein after its founders, was finalized. It is interesting to note that their theory unified all
known interactions at that time. As a time flew, more interactions were known and it became clear that to unify all of them, more extra dimensions are needed. At present, one of the promising theories to unify all interactions is M/string theory.

One of the distinguishing features of M/string theories is the presence of the curvature-squared corrections in the Lagrangian.
Scherk and Schwarz~\cite{sch-sch}  demonstrated the presence of the $R^2$ and
$R_{\mu \nu} R^{\mu \nu}$ terms in the Lagrangian of the Virasoro-Shapiro
model~\cite{VSh1, VSh2}; presence of the term of $R^{\mu \nu \lambda \rho}
R_{\mu \nu \lambda \rho}$ type was found in~\cite{Candelas_etal} for the low-energy limit
of the $E_8 \times E_8$ heterotic superstring theory~\cite{Gross_etal} to match the kinetic term
of the Yang-Mills field. Later it was demonstrated~\cite{zwiebach} that the only
combination of quadratic terms that leads to a ghost-free nontrivial gravitation
interaction is the Gauss-Bonnet (GB) term:

$$
L_{GB} = L_2 = R_{\mu \nu \lambda \rho} R^{\mu \nu \lambda \rho} - 4 R_{\mu \nu} R^{\mu \nu} + R^2.
$$

\noindent This term, first discovered
by Lanczos~\cite{Lanczos1, Lanczos2} (and so sometimes it is referred to
as the Lanczos term), is an Euler topological invariant in (3+1)-dimensional
space-time, but in (4+1) and higher dimensions it gives nontrivial contribution to the equations of motion.
Zumino~\cite{zumino} extended Zwiebach's result on
higher-than-squared curvature terms, supporting the idea that the low-energy limit of the unified
theory might have a Lagrangian density as a sum of contributions of different powers of curvature. The sum of all possible Euler topological invariants, which give nontrivial
contribution to the equations of motion in a particular number of space-time dimensions, form  more general Lovelock gravity~\cite{Lovelock}.

When one hears of the extra spatial dimensions, the natural question arises -- where they are? Our everyday experience clearly indicates there are three spatial dimensions, and
experiments in physics and theory support this (for instance, in Newtonian gravity in more then three space dimensions there are no stable orbits, while we clearly see they are).
The string theorists working with extra dimensions came with an answer -- the extra spatial dimensions are compact -- they are compactified on a very small scale, so small that we
cannot sense them with our level of equipment. But with that answer, another natural question comes to mind -- how come that they are compact? The answer to this question is not
that simple. One of
the ways to hide extra dimensions and to recover four-dimensional physics, is to consider so-called ``spontaneous compactification'' solution. Exact static solutions with the metric as a cross product of a
(3+1)-dimensional Minkowski space-time and a constant curvature ``inner space'',  were found for the first time in~\cite{add_1} (the generalization for
a constant curvature Lorentzian manifold was done in~\cite{Deruelle2}).
For cosmology, it is more useful to consider the four-dimensional part given by a Friedmann-Robertson-Walker metric, and the size of extra dimensions to be time-dependent rather
then static. In~\cite{add_4} it was demonstrated that in order to have a more realistic model one needs to consider the dynamical evolution of the extra-dimensional scale factor.
In~\cite{Deruelle2}, the equations of motion with time-dependent scale factors were written for arbitrary Lovelock order in the special case of a spatially flat metric (the results were further proven and extended in~\cite{prd09}).
The results of~\cite{Deruelle2} were further analyzed for the special case of 10 space-time dimensions in~\cite{add_10}.
In~\cite{add_8}, the dynamical compactification was studied with use of the Hamiltonian formalism.
More recently, searches for spontaneous  compactifications were made in~\cite{add13}, where
the dynamical compactification of the (5+1) Einstein-Gauss-Bonnet (EGB) model was considered; in \cite{MO04, MO14} with different metric {\it Ans\"atze} for scale factors
corresponding to (3+1)- and extra-dimensional parts. Also, apart from the
cosmology, the recent analysis has focused on
properties of black holes in Gauss-Bonnet~\cite{alpha_12, add_rec_1, add_rec_2, addn_1, addn_2} and Lovelock~\cite{add_rec_3, add_rec_4, addn_3, addn_4, addn_4.1} gravities, features of gravitational collapse in these
theories~\cite{addn_5, addn_6, addn_7}, general features of spherical-symmetric solutions~\cite{addn_8}, and many others.

If we want to find exact cosmological solutions, the most common {\it Ansatz} used for the scale factor is exponential or power law.
Exact solutions with exponents  for both  the (3+1)- and extra-dimensional scale factors were studied for the first time in~\cite{Is86}, and exponentially increasing (3+1)-dimensional
scale factor and an exponentially shrinking extra-dimensional scale factor were described.
Power-law solutions have been considered  in~\cite{Deruelle1, Deruelle2} and more  recently in~\cite{mpla09, prd09, Ivashchuk, prd10, grg10} so that by now there is an  almost complete description of the solutions of this kind
(see also~\cite{PT} for comments regarding physical branches of the power-law solutions).
Solutions with exponential scale factors~\cite{KPT} have also been studied in detail, namely, the models with both variable~\cite{CPT1} and constant~\cite{CST2} volume; the general scheme for
finding anisotropic exponential solutions in EGB was developed and generalized for general Lovelock gravity of any order and in any dimensions~\cite{CPT3}. The stability of the exponential solutions was addressed in~\cite{my15}
(see also~\cite{iv16} for stability of general exponential
solutions in EGB gravity), and it was
demonstrated that only a handful of the solutions found and described in~\cite{CPT3} could be called ``stable'', while the most of them are either unstable or have neutral/marginal stability.

In order to find all possible cosmological regimes in Einstein-Gauss-Bonnet gravity, one needs to go beyond an exponential or power-law {\it Ansatz} and keep the scale factor generic.
We are especially interested in models that allow dynamical compactification, so that we
consider the metric as the product of a spatially three-dimensional and extra-dimensional parts. In that case the three-dimensional part is ``our Universe'' and we expect for this part to expand
while the extra-dimensional part should be suppressed in size with respect to the three-dimensional one. In~\cite{CGP1} we demonstrated the there exist the phenomenologically
sensible regime when the curvature of the extra dimensions is negative and the Einstein-Gauss-Bonnet theory does not admit a maximally symmetric solution. In this case both the
three-dimensional Hubble parameter and the extra-dimensional scale factor asymptotically tend to the constant values. In~\cite{CGP2} we performed a detailed analysis of the cosmological dynamics in this model
with generic couplings. Recent analysis of this model~\cite{CGPT} revealed that, with an additional constraint on couplings, Friedmann-type late-time behavior
could be restored.

With the exponential and power-law solutions described in the mentioned above papers, another natural question arise -- could these solutions describe realistic compactification or
are they just solutions with no connection to the reality? To answer this question, we have considered the cosmological model in EGB gravity with the spatial part being the product
of three- and extra dimensional parts with both subspaces being spatially flat. As both subspaces are spatially flat, the equations of motion could be rewritten in terms of Hubble
parameters and then they become first order differential equations and could be analytically analyzed to find all possible regimes, asymptotes, exponential and power-law solutions.
For vacuum EGB model it was done in~\cite{my16a} and reanalyzed in~\cite{my18a}. The results suggest that in the vacuum model has two physically viable regimes -- first of them is the smooth transition from high-energy GB Kasner to low-energy GR Kasner. This regime
exists for $\alpha > 0$ (Gauss-Bonnet coupling) at $D=1,\,2$ (the number of extra dimensions) and for $\alpha < 0$ at $D \geqslant 2$ (so that at $D=2$ it appears for both signs of $\alpha$). Another viable regime is the smooth
transition from high-energy GB
Kasner to anisotropic exponential regime with expanding three-dimensional section (``our Universe'') and contracting extra dimensions; this regime occurs only for $\alpha > 0$ and at
$D \geqslant 2$.

The same analysis but for EGB model with $\Lambda$-term was performed in~\cite{my16b, my17a} and reanalyzed in~\cite{my18a}. The results suggest that the only realistic regime is the transition from high-energy GB Kasner to anisotropic exponential
solution, it requires $D \geqslant 2$, see~\cite{my16b, my17a, my18a} for exact limits on ($\alpha, \Lambda$).
The low-energy GR Kasner is
forbidden in the presence of the $\Lambda$-term so the corresponding transition do not occur.

In these studies we have made two important assumptions -- we considered both subspaces being isotropic and spatially flat. But what will happens in we lift these conditions?
Indeed, the spatial section
being a product of two isotropic spatially-flat subspaces could hardly be called ``natural'', so that we considered the effects of anisotropy and spatial curvature in~\cite{PT2017}. The initial anisotropy could affect the results greatly -- indeed, say, in vacuum $(4+1)$-dimensional EGB gravity with Bianchi-I-type metric (all the directions are independent) the only future asymptote is nonstandard singularity~\cite{prd10}. Our analysis~\cite{PT2017} suggest that the transition from Gauss-Bonnet Kasner
regime to anisotropic exponential expansion (with expanding
three and contracting extra dimensions) is stable with respect to breaking the symmetry within both three- and extra-dimensional subspaces. However, the details of the dynamics in
$D=2$ and $D \geqslant 3$ are different -- in the latter there exist anisotropic exponential solutions with ``wrong'' spatial splitting and all of them are accessible from generic
initial conditions. For instance, in $(6+1)$-dimensional space-time there are anisotropic exponential solutions with $[3+3]$ and $[4+2]$ spatial splittings, and some of the initial
conditions in the vicinity of $E_{3+3}$ actually end up in $E_{4+2}$ -- the exponential solution with four and two isotropic subspaces. In other words, generic initial conditions
could easily end up with ``wrong'' compactification, giving ``wrong'' number of expanding spatial dimensions (see~\cite{PT2017} for details).

The effect of the spatial curvature on the cosmological dynamics could be dramatic -- say, positive curvature changes the inflationary asymptotics~\cite{infl1, infl2}. In the case
of EGB gravity the influence of the spatial curvature
reveal itself only if the curvature of the extra dimensions is negative and $D \geqslant 3$
-- in that case there exist  ``geometric frustration'' regime, described in~\cite{CGP1, CGP2} and further investigated in~\cite{CGPT}.

The current manuscript could be called a spiritual successor of~\cite{my16a, my16b, my17a, my18a} -- now we are performing the same analysis but for cubic Lovelock gravity. In this paper we consider only $D=3, 4$ (the number of the extra spatial dimensions) for vacuum case, other $D$ cases, as well as
$\Lambda$-term case and possible influence of anisotropy, spatial curvature and different kinds of matter source are to be considered in
the papers to follow.

The manuscript is structured as follows: first we introduce Lovelock gravity and derive the equations of motion in the general form for the spatially-flat (Bianchi-I-type) metrics. Then we add our {\it Ansatz} and write down
simplified equations. After that we describe the scheme we are going to use to analyze the particular cases. Then we consider particular cases with $D=3$ and $D=4$; for the former, we are going to describe the scheme step-by-step.
In each section, dedicated to the particular case, we describe it and briefly summarize its features. Finally we summarize both cases,  discuss their differences and similarities. After that we compare the dynamics in this cubic Lovelock with the dynamics in quadratic Lovelock (Einstein-Gauss-Bonnet) case, described in~\cite{my16a, my18a}. At last, we draw conclusions and formulate
perspective directions for further investigations.

\section{Equations of motion}

Lovelock gravity~\cite{Lovelock} has the following structure: its Lagrangian is constructed from terms

\begin{equation}
L_n = \frac{1}{2^n}\delta^{i_1 i_2 \dots i_{2n}}_{j_1 j_2 \dots
j_{2n}} R^{j_1 j_2}_{i_1 i_2}
 \dots R^{j_{2n-1} j_{2n}}_{i_{2n-1} i_{2n}}, \label{lov_lagr}
\end{equation}

\noindent where $\delta^{i_1 i_2 \dots i_{2n}}_{j_1 j_2 \dots
j_{2n}}$ is the generalized Kronecker delta of the order $2n$.
One can verify that $L_n$ is Euler invariant in $D < 2n$ spatial dimensions and so it would not give nontrivial contribution into the equations of motion. So that the
Lagrangian density for any given $D$ spatial dimensions is sum of all Lovelock invariants (\ref{lov_lagr}) upto $n=\[\dac{D}{2}\]$ which give nontrivial contributions
into equations of motion:

\begin{equation}
{\cal L}= \sqrt{-g} \sum_n c_n L_n, \label{lagr}
\end{equation}

\noindent where $g$ is the determinant of metric tensor,
$c_n$ are coupling constants of the order of Planck length in $2n$
dimensions and summation over all $n$ in consideration is assumed. The metric {\it ansatz} has the form

\begin{equation}\label{metric}
g_{\mu\nu} = \diag\{ -1, a_1^2(t), a_2^2(t),\ldots, a_n^2(t)\}.
\end{equation}

\noindent As we mentioned earlier, we are interested in the dynamics in cubic Lovelock gravity, so we consider $n$ up to three ($n=0$ is boundary term while $n=1$ is Einstein-Hilbert, $n=2$ is Gauss-Bonnet and $n=3$ is cubic
Lovelock contributions).
Substituting metric (\ref{metric}) into the Lagrangian and following the usual procedure gives us the equations of motion:

\begin{equation}
\begin{array}{l}
2 \[ \sum\limits_{j\ne i} (\dot H_j + H_j^2)
+ \sum\limits_{\substack{\{ k > l\} \\ \ne i}} H_k H_l \] + 8\alpha \[ \sum\limits_{j\ne i} (\dot H_j + H_j^2) \sum\limits_{\substack{\{k>l\} \\ \ne \{i, j\}}} H_k H_l +
3 \sum\limits_{\substack{\{ k > l >  \\   m > n\} \ne i}} H_k H_l H_m H_n \] + \\ \\
+ 144\beta\[ \sum\limits_{j\ne i} (\dot H_j + H_j^2) \sum\limits_{\substack{\{k>l>m> \\ n\} \ne \{i, j\}}} H_k H_l H_m H_n + 5 \sum\limits_{\substack{\{ k > l > m >  \\   n > p > q\} \ne i}} H_k H_l H_m H_n H_p H_q   \]
- \Lambda = 0
\end{array} \label{dyn_gen}
\end{equation}

\noindent as the $i$th dynamical equation. The first Lovelock term---the Einstein-Hilbert contribution---is in the first set of brackets, the second term---Gauss-Bonnet---is in the second set and the third -- cubic Lovelock
term---is in the third set; $\alpha$
is the coupling constant for the Gauss-Bonnet contribution while $\beta$ is the coupling constant for cubic Lovelock; we put the corresponding constant for Einstein-Hilbert contribution to unity.
Also, since in this section we consider spatially flat cosmological models, scale
factors do not hold much in the physical sense and the equations are rewritten in terms of the Hubble parameters $H_i = \dot a_i(t)/a_i(t)$. Apart from the dynamical equations, we write down the constraint equation

\begin{equation}
\begin{array}{l}
2 \sum\limits_{i > j} H_i H_j + 24\alpha \sum\limits_{\substack{i > j >\\  k > l}} H_i H_j H_k H_l + 720\beta \sum\limits_{\substack{i > j > k \\ > l> m > n}} H_i H_j H_k H_l H_m H_n= \Lambda.
\end{array} \label{con_gen}
\end{equation}

As mentioned in the Introduction,
we want to investigate the particular case with the scale factors split into two parts -- separately three dimensions (three-dimensional isotropic subspace), which are supposed to represent our world, and the remaining represent the extra dimensions ($D$-dimensional isotropic subspace). So we put $H_1 = H_2 = H_3 = H$ and $H_4 = \ldots = H_{D+3} = h$ ($D$ designs the number of additional dimensions) and the
equations take the following form: the
dynamical equation that corresponds to $H$,

\begin{equation}
\begin{array}{l}
2 \[ 2 \dot H + 3H^2 + D\dot h + \dac{D(D+1)}{2} h^2 + 2DHh\] + 8\alpha \[ 2\dot H \(DHh + \dac{D(D-1)}{2}h^2 \) + \right. \\
 \\ \left. + D\dot h \(H^2 + 2(D-1)Hh + \dac{(D-1)(D-2)}{2}h^2 \) +
2DH^3h + \dac{D(5D-3)}{2} H^2h^2 + \right. \\
\\ \left. + D^2(D-1) Hh^3 + \dac{(D+1)D(D-1)(D-2)}{8} h^4 \] +  \\ \\
+  144\beta \[ \dot H \(Hh^3 \dac{D(D-1)(D-2)}{3} +
h^4\dac{D(D-1)(D-2)(D-3)}{12} \) + \right. \\ \\
+ \left. D \dot h
\(H^2 h^2 \dac{(D-1)(D-2)}{2} +  Hh^3 \dac{(D-1)(D-2)(D-3)}{3} + \right. \right. \\
\\ + \left. \left.  h^4\dac{(D-1)(D-2)(D-3)(D-4)}{24} \) +  H^3h^3 \dac{D(D-1)(D-2)}{3} + \right. \\ \\
+ \left. H^2h^4 \dac{D(D-1)(D-2)(7D-9)}{24} +  Hh^5 \dac{D^2(D-1)(D-2)(D-3)}{12} + \right. \\ \\
+ \left. h^6 \dac{(D+1)D(D-1)(D-2)(D-3)(D-4)}{144} \]
 - \Lambda=0,
\end{array} \label{H_gen}
\end{equation}

\noindent the dynamical equation that corresponds to $h$,

\begin{equation}
\begin{array}{l}
2 \[ 3 \dot H + 6H^2 + (D-1)\dot h + \dac{D(D-1)}{2} h^2 + 3(D-1)Hh\] + 8\alpha \[ 3\dot H \(H^2 + \right. \right. \\
\\ \left. \left. + 2(D-1)Hh +  \dac{(D-1)(D-2)}{2}h^2 \) +  (D-1)\dot h \(3H^2 + 3(D-2)Hh + \right. \right. \\
\\  \left. \left. +
\dac{(D-2)(D-3)}{2}h^2 \) + 3H^4 +  9(D-1)H^3h + 3(D-1)(2D-3) H^2h^2 + \right. \\
\\ \left. + \dac{3(D-1)^2 (D-2)}{2} Hh^3 +   \dac{D(D-1)(D-2)(D-3)}{8} h^4 \] + \\ \\  + 144\beta\[ \dot H \( H^2 h^2 \dac{3(D-1)(D-2)}{2} +  Hh^3(D-1)(D-2)(D-3) + \right. \right. \\
\\ + \left. \left.
h^4 \dac{(D-1)(D-2)(D-3)(D-4)}{8} \)  + (D-1)\dot h \(  H^3 h (D-2) + \right. \right. \\
\\ + \left. \left. H^2h^2 \dac{3(D-2)(D-3)}{2} +  Hh^3 \dac{(D-2)(D-3)(D-4)}{2} + \right. \right. \\
\\  \left. \left. +
h^4 \dac{(D-2)(D-3)(D-4)(D-5)}{24}  \) + H^4 h^2 \dac{3(D-1)(D-2)}{2} + \right. \\ \\ \left. + H^3h^3 \dac{(D-1)(D-2)(11D-27)}{6} +  H^2h^4 \dac{3(D-1)(D-2)^2(D-3)}{4} +
\right. \\ \\ \left. + Hh^5 \dac{(D+1)(D-1)(D-2)(D-3)(D-4)}{12} + \right. \\ \\ \left. +
 h^6\dac{D(D-1)(D-2)(D-3)(D-4)(D-5)}{144} \]
- \Lambda =0,
\end{array} \label{h_gen}
\end{equation}

\noindent and the constraint equation,

\begin{equation}
\begin{array}{l}
2 \[ 3H^2 + 3DHh + \dac{D(D-1)}{2} h^2 \] + 24\alpha \[ DH^3h + \dac{3D(D-1)}{2}H^2h^2 + \right. \\ \\ \left. + \dac{D(D-1)(D-2)}{2}Hh^3 +  \dac{D(D-1)(D-2)(D-3)}{24}h^4\] +
 720\beta \[ H^3 h^3 \dac{D(D-1)(D-2)}{6} + \right. \\ \\ \left. +
 H^2 h^4 \dac{D(D-1)(D-2)(D-3)}{8} +  Hh^5 \dac{D(D-1)(D-2)(D-3)(D-4)}{40} + \right. \\ \\ \left. + h^6 \dac{D(D-1)(D-2)(D-3)(D-4)(D-5)}{720}    \] = \Lambda.
\end{array} \label{con2_gen}
\end{equation}

Looking at (\ref{H_gen})--(\ref{con2_gen}) one can notice that the structure of the equations depends on the number of extra dimensions $D$ (terms with $(D-4)$ multiplier nullifies in $D=4$ and so on).
In previous papers, dedicated to study cosmological dynamics in EGB gravity, we
performed analysis in all dimensions, sensitive to EGB case~\cite{my16a, my16b, my17a, my18a}. In the cubic Lovelock, the structure of the equations of motion is different in $D=3, 4, 5$ and in the general $D \geqslant 6$ cases.
Also, since the current paper is dedicated to the vacuum case, we have $\Lambda \equiv 0$.

\section{General scheme}

The procedure of the analysis is exactly the same as described in our pervious papers~\cite{my16a, my16b, my17a, my18a} and is as follows:

\begin{itemize}

 \item we solve (\ref{con2_gen}) with respect to $H$ -- one can see that (\ref{con2_gen}) is cubic with respect to $H$ and sixth order with respect to $h$, so that to have analytical solutions, we solve it for $H$;
 as a result we have three branches $H_1$, $H_2$ and $H_3$. In lower-dimensional cases we wrote down solutions explicitly, but in higher dimensions they become quite bulky, so draw $H(h)$ curves instead. If we take the discriminant of (\ref{con2_gen}) with respect to $H$, and then its discriminant with respect to $h$, we obtain critical values for $(\alpha, \beta)$ which separate qualitatively different cases;

\item we find analitically isotropic exponential solutions: to do this we substitute $\dot H = \dot h \equiv 0$ as well as $h = H$ into (\ref{H_gen})--(\ref{con2_gen}); the system simplifies into a single equation
and we solve it, finding not only roots but also the ranges of $(\alpha, \beta)$ where they exist;

\item we find analitically anisotropic exponential solutions: to do this we substitute \linebreak \mbox{$\dot H = \dot h \equiv 0$} into (\ref{H_gen})--(\ref{con2_gen}); the system could be brought to two equations: bi-six order polynomial
in $h$ with powers of $\alpha$ and $\beta$ as coefficients and
$H = H(h, \alpha, \beta)$. Both of them are usually higher-order with respect to their arguments so retrieving the solutions in radicand is impossible. But if we consider the discriminant of the former of them, the resulting
equation gives us critical values for $(\alpha, \beta)$ which separate areas with different number of roots;

\item altogether first three items provides us with a set of critical values for $(\alpha, \beta)$ which separate domains with different dynamics;

 \item we solve (\ref{H_gen})--(\ref{h_gen}) with respect to $\dot H$ and $\dot h$;

 \item we substitute obtained $H_i$ curves into $\dot H$ and $\dot h$ and obtain the latter as a single-variable functions: $\dot H(h)$ and $\dot h(h)$;

 \item the obtained $\dot H(h)$ and $\dot h(h)$ expressions are analyzed for all possible domains in $(\alpha, \beta)$ space to obtain all possible regimes;

 \item obtained exponential regimes are compared with exact isotropic and anisotropic solutions (see~\cite{CPT3}) to find the nature of the exponential regimes in question;

 \item power-law regimes are analyzed in terms of Kasner exponents ($p_i = - H_i^2/\dot H_i$) to verify that low-energy power-law regimes are standard Kasner regimes with
 \mbox{$\sum p_i = \sum p_i^2 = 1$} while high-energy
 power-law regimes are Lovelock Kasner regimes with \linebreak \mbox{$\sum p_i = (2n-1) = 5$}.

\end{itemize}

The described above scheme allows us to completely describe all existing regimes for a given set of the parameters $(\alpha, \beta)$. In the first case to consider, $D=3$, we are going to completely describe all steps in
detail.

Before continuing with the particular cases, it is necessary to introduce the notations we are going to use through the paper. We denote Kasner regime as $K_i$ where $i$ is the
total expansion rate in terms of the Kasner exponents $\sum p_i = (2n-1)$ where $n$ is the corresponding order of the Lovelock term (see, e.g.,~\cite{prd09}).
This way, for Einstein-Hilbert it is $n=1$ and so $\sum p_i = 1$ (see~\cite{kasner}) and the corresponding regime is $K_1$, which is usual low-energy regime in vacuum
EGB case (see~\cite{my17a, my18a}) and we expect it to play the same role here. For Gauss-Bonnet it is $n=2$ and so $\sum p_i = 3$ and the regime $K_3$ is typical high-energy regime for vacuum EGB case (again, see~\cite{my17a, my18a}). Finally, for cubic Lovelock it is $n=3$ and so $\sum p_i = 5$ and the regime $K_5$ is expected to be typical high-energy regime for
this case.

Another power-law regime is what we call ``generalized Taub'' (see~\cite{Taub} for the original solution).
It is the regime which was mistakenly taken for $K_3$ in~\cite{my17a}, but in~\cite{my18a} it was
corrected and explained (they both have $\sum p_i = 3$ which causes misinterpreting). It is a situation when for one of the subspaces the Kasner exponent $p \equiv - H^2/\dot H$ is equal to zero and for another --
to unity. So we denote
$P_{1, 0}$ the case with $p_H = 1, p_h = 0$ and $P_{0, 1}$ the case with $p_H = 0, p_h = 1$.

We denote the exponential solutions as $E$ with subindex indicating its details -- $E_{iso}$ is isotropic exponential solution and $E_{3+D}$ is anisotropic -- with different Hubble
parameters corresponding to three- and extra-dimensional subspaces. In practice, in each particular case there are several different anisotropic exponential solutions, so that instead
of using $E_{3+D}$ we use $E_i$ where $i$ counts the number of the exponential solution ($E_1$, $E_2$ etc). In case if there are several isotropic exponential solutions, we count them with upper index:
$E_{iso}^1$, $E_{iso}^2$ etc.

The final regime is what we call ``nonstandard singularity'' and we denote is as $nS$.
It is the situation which arise in Lovelock gravity due to its nonlinear nature. Since the equations
(\ref{H_gen})--(\ref{h_gen}) are nonlinear with respect to the highest derivative ($\dot H$ and $\dot h$ in our case), when we solve them, the resulting expressions are ratios with
polynomials in both numerator and denominator. So there exist a situation then the denominator is equal to zero for finite values of $H$ and\/or $h$. This situation is singular, as the
curvature invariants diverges, but it happening for finite values of $H$ and\/or $h$. Tipler~\cite{Tipler} call this kind of singularity as ``weak'' while Kitaura and
Wheeler~\cite{KW1, KW2} -- as ``type II''. Our previous research demonstrate that this kind of singularity is wide spread in EGB cosmology -- in particular, in totally anisotropic
(Bianchi-I-type) $(4+1)$-dimensional vacuum cosmological model it is the only future asymptote~\cite{prd10}.

\section{$D=3$ case}

In this case the equations of motion (\ref{H_gen})--(\ref{con2_gen}) take form ($H$-equation, $h$-equation, and constraint correspondingly)

\begin{equation}
\begin{array}{l}
4\dot H + 6H^2 + 6\dot h + 12h^2 + 12Hh + 8\alpha \( 6\dot Hh (H + h) + 3\dot h (H^2 + h^2 + 4Hh) +18H^2h^2 + \right.\\ \left. +18Hh^3 + 3h^4 + 6H^3h\) + 144\beta \( 2 (\dot H + H^2)Hh^3 + 3(\dot h+ h^2)H^2h^2   \)
 = 0,
\end{array} \label{D3_H}
\end{equation}

\begin{equation}
\begin{array}{l}
6\dot H + 12H^2 + 4\dot h + 6h^2 + 12Hh + 8\alpha \( 3\dot H (H^2 + 4Hh + h^2) + 6\dot h H (H+h) + 6Hh^3 + \right. \\ \left. + 18H^2h^2 + 18H^3h + 3H^4  \) + 144\beta \( 3(\dot H + H^2) H^2h^2 + 2(\dot h + h^2)H^3h   \)
 = 0,
\end{array} \label{D3_h}
\end{equation}

\begin{equation}
\begin{array}{l}
6 H^2 + 18Hh + 6h^2 + 24\alpha (3H^3h + 9H^2h^2 + 3Hh^3 ) + 720\beta H^3h^3 = 0.
\end{array} \label{D3_con}
\end{equation}

As mentioned above, for this case we describe each step with details while in the cases to
follow we skip the details.  Finding the discriminant of the discriminant of (\ref{D3_con}) with respect to $H$ gives us two critical values for
$\mu \equiv \beta/\alpha^2$; $\mu_1 = 1/6$, $\mu_2=3/2$. These are two values for $\mu$ which qualitatively change the behavior of the $H(h)$ curves.

To find out the details of isotropic exponential solutions, let us substitute $\dot H = \dot h \equiv 0$ and $h=H$ into (\ref{D3_H})--(\ref{D3_con}); the system simplifies to a single equation

\begin{equation}
\begin{array}{l}
720\beta H^6 + 360\alpha H^4 + 30 H^2 = 0,
\end{array} \label{D3_iso1}
\end{equation}

\noindent and it has trivial solution $H=0$ as well as up to two more solutions

\begin{equation}
\begin{array}{l}
H^2 = - \dac{3\alpha \pm \sqrt{9\alpha^2 - 6\beta}}{12\beta}.
\end{array} \label{D3_iso2}
\end{equation}

\noindent Analyzing (\ref{D3_iso2}) leads us to the following: we have one isotropic exponential solution if ($\alpha < 0$, $\beta < 0$) and ($\alpha > 0$, $\beta < 0$), and two
solutions if $\alpha < 0$, $\beta > 0$, $0 < \mu = \beta/\alpha^2 < 3/2$; in all other cases (\ref{D3_iso2}) is imagenary and so the isotropic solutions are absent.

As a next step we find out when anisotropic exponential solutions exist; to do this we substitute $\dot H = \dot h \equiv 0$ (but $h\ne H$) into (\ref{D3_H})--(\ref{D3_con}); the resulting equations could be solved to
obtain $h$ and $H$:

\begin{equation}
\begin{array}{l}
147456\mu^2\zeta^6 + (93312\mu^3 - 96768\mu^2 - 32256\mu)\zeta^5 + (21504\mu + 2304 - 1908\mu^2)\zeta^4 + \\ + (1056\mu - 1920 - 288\mu^2)\zeta^3 + (304-192\mu)\zeta^2 + (18\mu -24)\zeta + 1 = 0;  \\
H = - 8 h \dac{P_1}{P_2},\quad\mbox{where}~\zeta=\alpha h^2,~\mu=\beta/\alpha^2,
\end{array} \label{D3_aniso1}
\end{equation}

\noindent and $P_1$ and $P_2$ are bulky polynomials up to $\zeta^5$ and $\mu^4$ orders. The discriminant of (\ref{D3_aniso1}) is 18th-order polynomial in $\mu$ and have roots: single root $\mu_1 = 1/6$, quadruple roots
\mbox{$\mu_3 = (-2\sqrt[3]{100}/27 + 14/27) \approx 0.175$ and $\mu_4=2/3$}, and single root $\mu_2=3/2$.
%As it is widely known, roots of the even powers do not change the sign of the expression, so only $\mu_{1, 2}$ matters here, and the
%values exactly coincide with those obtained from $H(h)$ analysis.
So that for $\alpha < 0$ we have: for $\mu < 1/6$ (including $\mu < 0$) we have $\zeta > 0$, so that $h^2 < 0$ and so no real solutions for $h$; for
$3/2 > \mu > 1/6$ there are
no real solutions for $\zeta$; at $\mu = 3/2$ we have double root for $\zeta < 0$ and so $h = \pm \sqrt{\alpha \zeta}$, and finally for $\mu > 3/2$ we have two distinct roots $\zeta_{1, 2} < 0$ and so
$h = \pm \sqrt{\alpha \zeta_{1, 2}}$. To summarize, for $\alpha < 0$ we have exponential solutions only for $\mu \geqslant 3/2$.

For $\alpha > 0$ the situation is the following: for $\mu \leqslant 1/6$ (including $\mu < 0$) there are two distinct roots $\zeta_{1, 2} < 0$ and so $h = \pm \sqrt{\alpha \zeta_{1, 2}}$; for $\mu = 1/6$ additional
double root is added to the above roots (making total three different roots), for $\mu_3 > \mu > 1/6$ the double root from before is splitted making four different roots, at $\mu = \mu_3$ pairs of roots coinside leaving only
two different roots and for $\mu > \mu_3$ roots degrade into nonstandard singularities leaving us with no roots at all. At $\mu \geqslant 3/2$ we have roots again, but they are negative, so that the roots for
$h = \pm \sqrt{\alpha \zeta}$ are imagenary. Concluding, for $\alpha > 0$ we have exponential solutions only for $\mu \leqslant \mu_3$.

\begin{figure}
\centering
\includegraphics[width=0.7\textwidth, angle=0]{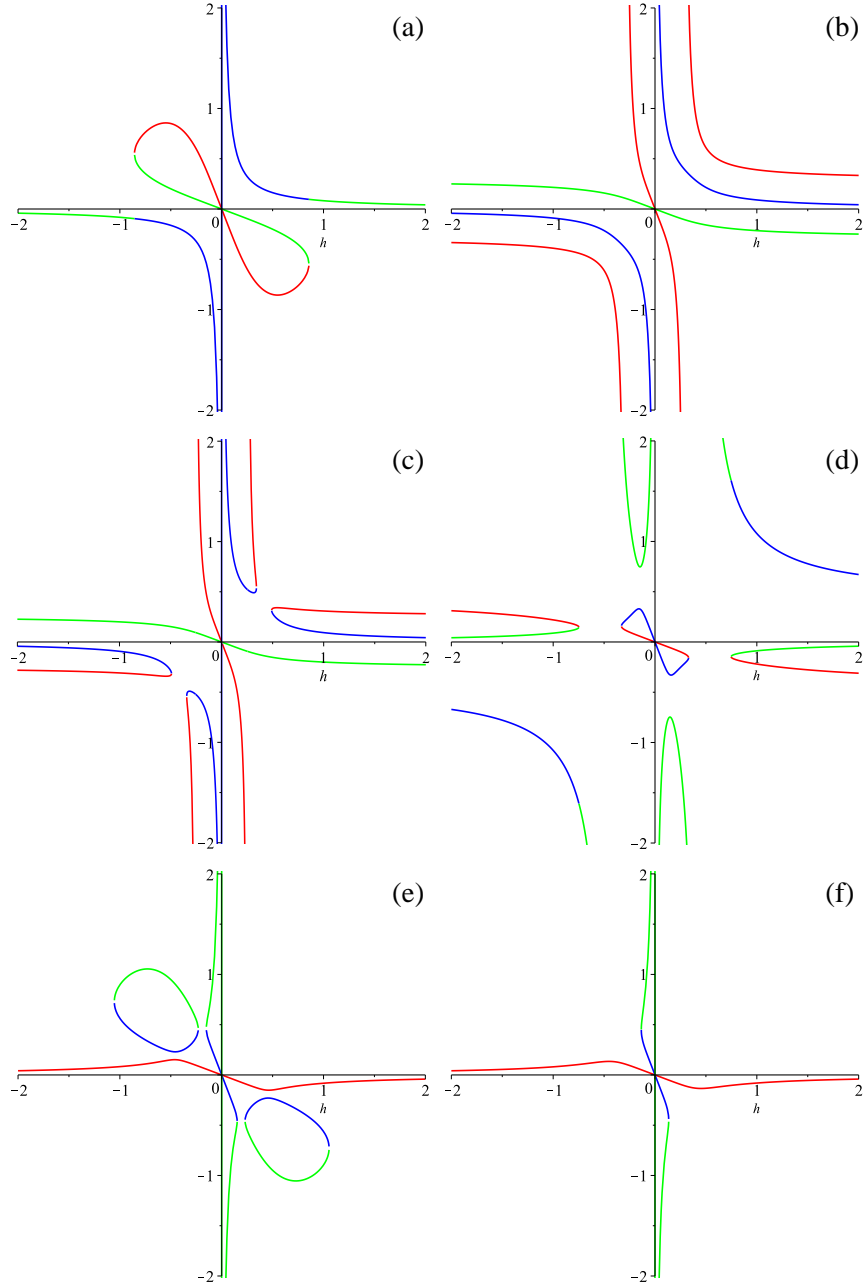}
\caption{$H(h)$ graphs for vacuum $D=3$ case: three different colors correspond to three different branches $H_1$, $H_2$ and $H_3$; $\alpha < 0$, $\beta < 0$ on (a) panel, $\alpha < 0$, $\beta > 0$, $\mu < 3/2$ on (b) panel, $\alpha < 0$, $\beta > 0$, $\mu > 3/2$ on (c) panel, $\alpha > 0$, $\beta < 0$ on
(d) panel, $\alpha > 0$, $\beta > 0$, $\mu < 0.3$ on (e) panel, and $\alpha > 0$, $\beta > 0$, $\mu > 0.3$ on (f) panel
(see the text for more details).}\label{D3_1}
\end{figure}

Now we can solve (\ref{D3_con}) with respect to $H$ and plot the resulting curves in Fig. \ref{D3_1}. There red curve corresponds to $H_1$, blue to $H_2$ and green
to $H_3$. The panels layout is as follows: $\alpha < 0$, $\beta < 0$ on (a) panel, $\alpha < 0$, $\beta > 0$, $\mu < 3/2$ on (b) panel, $\alpha < 0$, $\beta > 0$, $\mu > 3/2$ on (c) panel, $\alpha > 0$, $\beta < 0$ on
(d) panel, $\alpha > 0$, $\beta > 0$, $\mu < 0.3$ on (e) panel, and $\alpha > 0$, $\beta > 0$, $\mu > 0.3$ on (f) panel.
One can see that on (a) and (d) panels (and so $\beta < 0$ and arbitrary $\alpha$) we have one isotropic exponential solution (there exist $H=h$ solution) while on (b) panel ($\alpha < 0$, $\beta > 0$, $\eta > 2/3$) we have
two different isotropic exponential solutions; in all other cases there are no isotropic solutions.

\begin{figure}
\centering
\includegraphics[width=0.96\textwidth, angle=0]{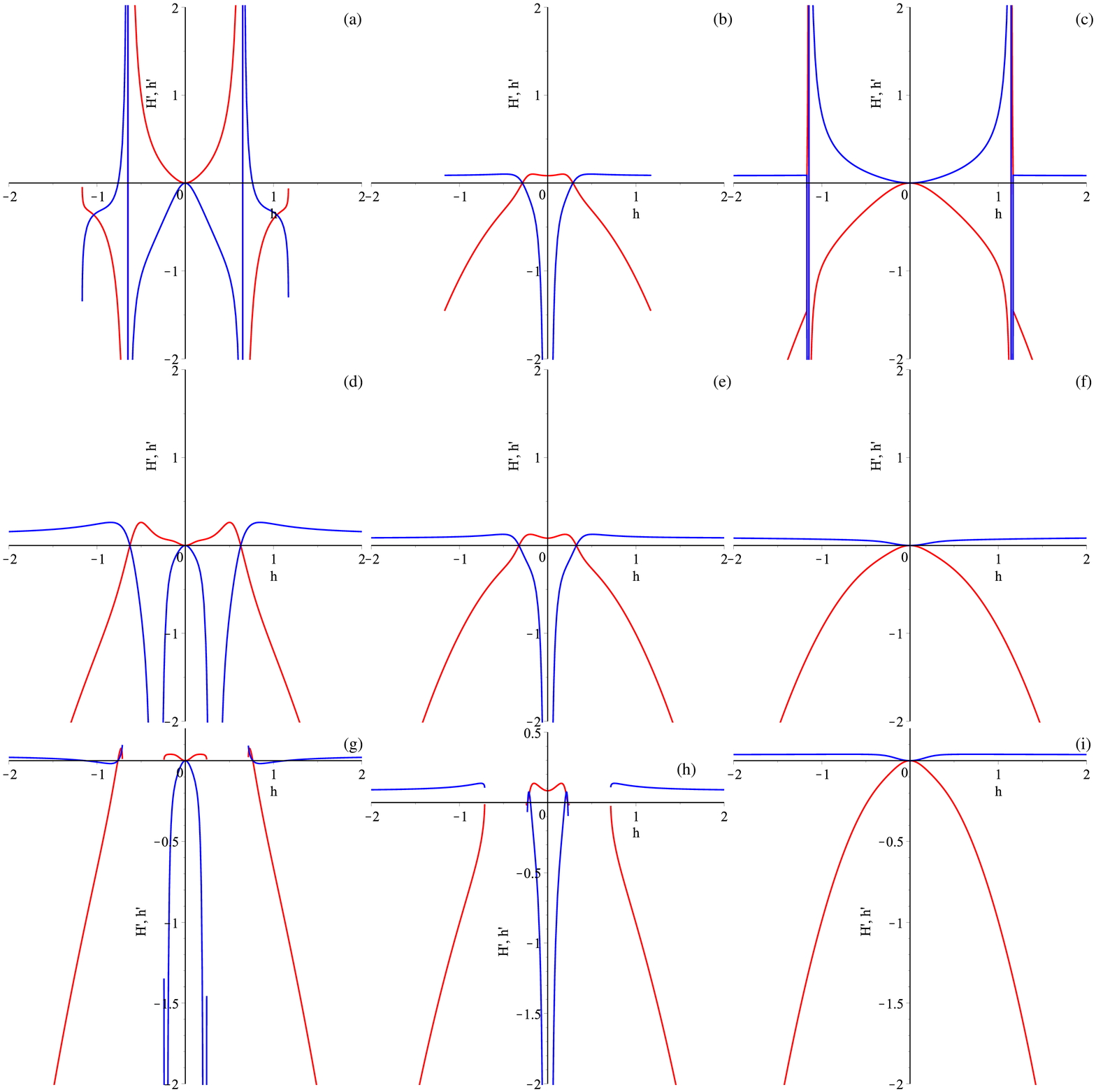}
\caption{$\dot H(h)$ and $\dot h(h)$ graphs for vacuum $D=3$ case:  $\dot h(h)$ in red and $\dot H(h)$ in blue;
$\alpha < 0$, $\beta < 0$ on (a)--(c) panels: $H_1$ branch on (a) panel, $H_2$ branch on (b) panel and $H_3$ branch on (c) panel;
$\alpha < 0$, $\beta > 0$, $\mu < 3/2$ on (d)--(f) panels: $H_1$ branch on (d) panel, $H_2$ branch on (e) panel and $H_3$ branch on (f) panel;
$\alpha < 0$, $\beta > 0$, $\mu > 3/2$ on (g)--(i) panels: $H_1$ branch on (g) panel, $H_2$ branch on (h) panel and $H_3$ branch on (i) panel;
%$\alpha > 0$, $\beta < 0$ on (d) panel, $\alpha > 0$, $\beta > 0$, $\mu < 0.3$ on (e) panel, and $\alpha > 0$, $\beta > 0$, $\mu > 0.3$ on (f) panel
(see the text for more details).}\label{D3_dHdh_1}
\end{figure}

\begin{figure}
\centering
\includegraphics[width=0.96\textwidth, angle=0]{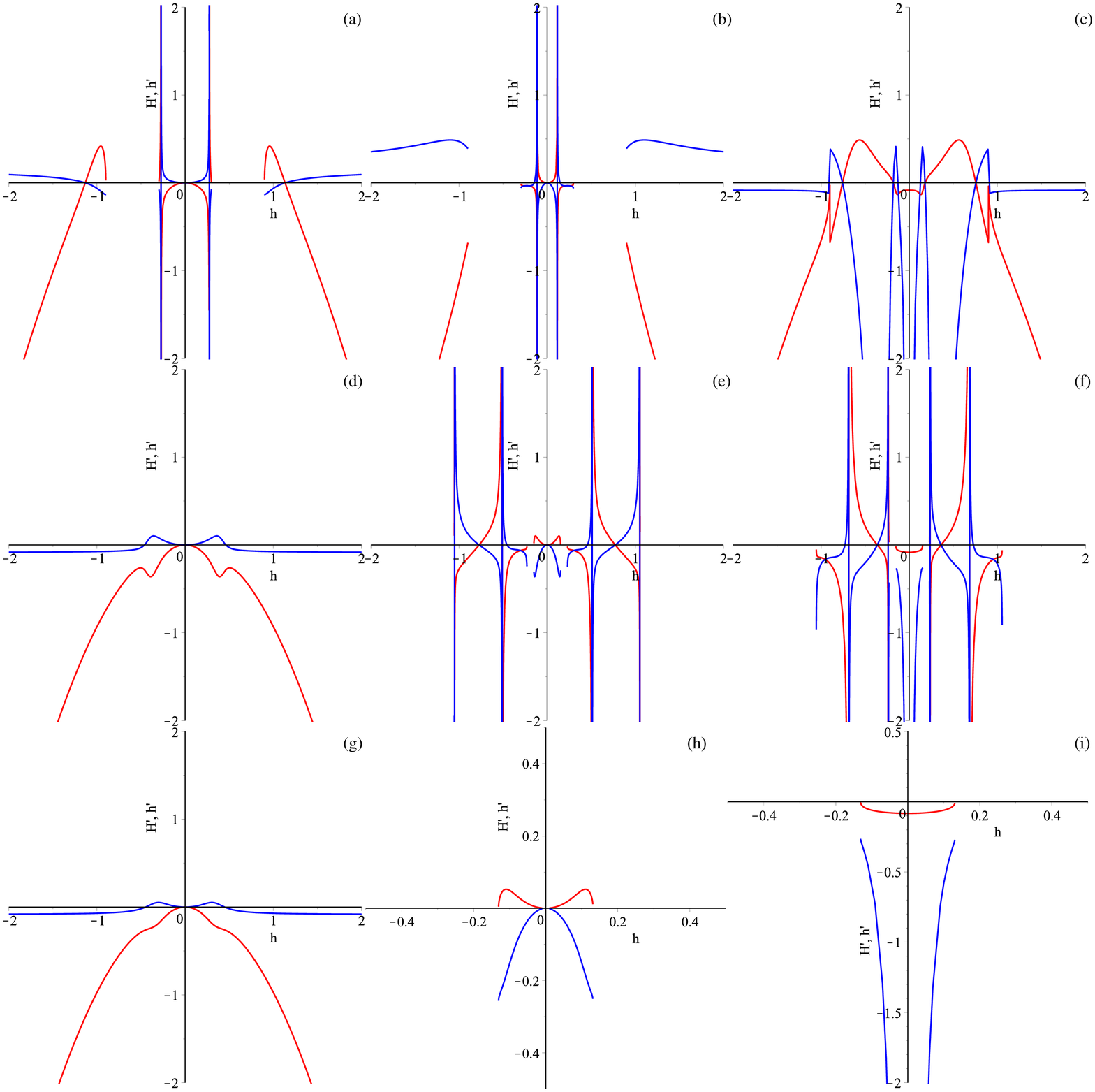}
\caption{$\dot H(h)$ and $\dot h(h)$ graphs for vacuum $D=3$ case:  $\dot h(h)$ in red and $\dot H(h)$ in blue;
$\alpha > 0$, $\beta < 0$ on (a)--(c) panels: $H_1$ branch on (a) panel, $H_2$ branch on (b) panel and $H_3$ branch on (c) panel;
$\alpha > 0$, $\beta > 0$, $\mu < 0.3$ on (d)--(f) panels: $H_1$ branch on (d) panel, $H_2$ branch on (e) panel and $H_3$ branch on (f) panel;
$\alpha > 0$, $\beta > 0$, $\mu > 0.3$ on (g)--(i) panels: $H_1$ branch on (g) panel, $H_2$ branch on (h) panel and $H_3$ branch on (i) panel;
%$\alpha > 0$, $\beta < 0$ on (d) panel, $\alpha > 0$, $\beta > 0$, $\mu < 0.3$ on (e) panel, and $\alpha > 0$, $\beta > 0$, $\mu > 0.3$ on (f) panel
(see the text for more details).}\label{D3_dHdh_2}
\end{figure}

The next step is derive $\dot H$ and $\dot h$ -- we solve (\ref{D3_H})--(\ref{D3_h}) with respect to them and substitute $H_1$, $H_2$ and $H_3$ branches separately. The resulting expressions are quite bulky so that we do not
write them down, but provide the graphs in Figs.~\ref{D3_dHdh_1}--\ref{D3_dHdh_2}. There we  presented $\dot h(h)$ as red and $\dot H(h)$ as blue curves and the panels layout is as follows: in Fig.~\ref{D3_dHdh_1} we
presented $\alpha < 0$ cases: $\beta < 0$ on (a)--(c) panels ($H_1$ branch on (a) panel, $H_2$ branch on (b) panel and $H_3$ branch on (c) panel),
$\beta > 0$, $\mu < 3/2$ on (d)--(f) panels ($H_1$ branch on (d) panel, $H_2$ branch on (e) panel and $H_3$ branch on (f) panel) and
$\beta > 0$, $\mu > 3/2$ on (g)--(i) panels ($H_1$ branch on (g) panel, $H_2$ branch on (h) panel and $H_3$ branch on (i) panel). In Fig.~\ref{D3_dHdh_2} we presented $\alpha > 0$ cases:
$\beta < 0$ on (a)--(c) panels ($H_1$ branch on (a) panel, $H_2$ branch on (b) panel and $H_3$ branch on (c) panel),
$\beta > 0$, $\mu < 0.3$ on (d)--(f) panels ($H_1$ branch on (d) panel, $H_2$ branch on (e) panel and $H_3$ branch on (f) panel) and
$\beta > 0$, $\mu > 0.3$ on (g)--(i) panels ($H_1$ branch on (g) panel, $H_2$ branch on (h) panel and $H_3$ branch on (i) panel).

\begin{figure}
\centering
\includegraphics[width=0.96\textwidth, angle=0]{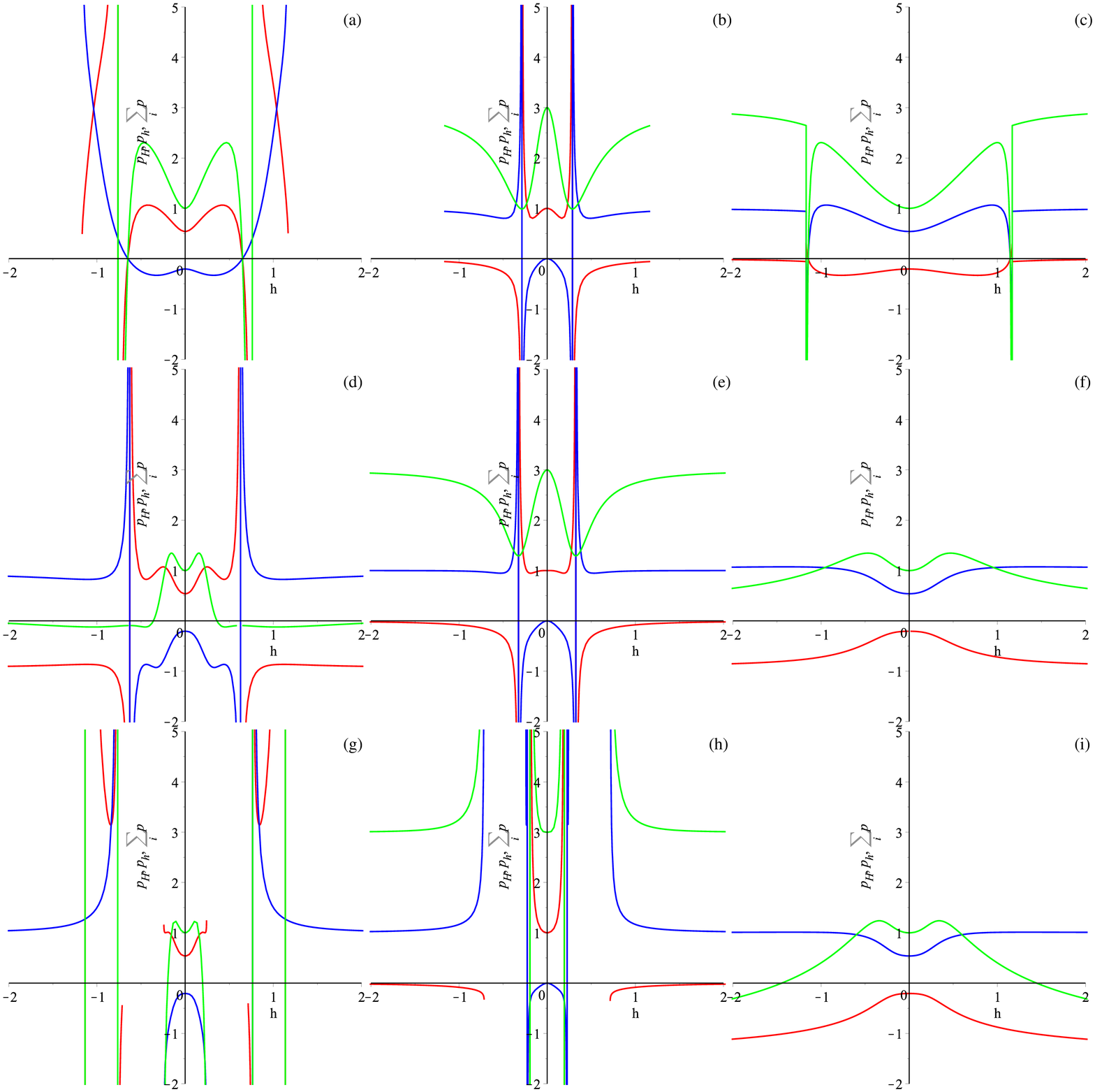}
\caption{$p_H$, $p_h$ and $\sum p_i$ graphs for vacuum $D=3$ case:  $p_H$ in red, $p_h$ in blue and $\sum p_i$ in green;
$\alpha < 0$, $\beta < 0$ on (a)--(c) panels: $H_1$ branch on (a) panel, $H_2$ branch on (b) panel and $H_3$ branch on (c) panel;
$\alpha < 0$, $\beta > 0$, $\mu < 3/2$ on (d)--(f) panels: $H_1$ branch on (d) panel, $H_2$ branch on (e) panel and $H_3$ branch on (f) panel;
$\alpha < 0$, $\beta > 0$, $\mu > 3/2$ on (g)--(i) panels: $H_1$ branch on (g) panel, $H_2$ branch on (h) panel and $H_3$ branch on (i) panel;
%$\alpha > 0$, $\beta < 0$ on (d) panel, $\alpha > 0$, $\beta > 0$, $\mu < 0.3$ on (e) panel, and $\alpha > 0$, $\beta > 0$, $\mu > 0.3$ on (f) panel
(see the text for more details).}\label{D3_ph_1}
\end{figure}

\begin{figure}
\centering
\includegraphics[width=0.96\textwidth, angle=0]{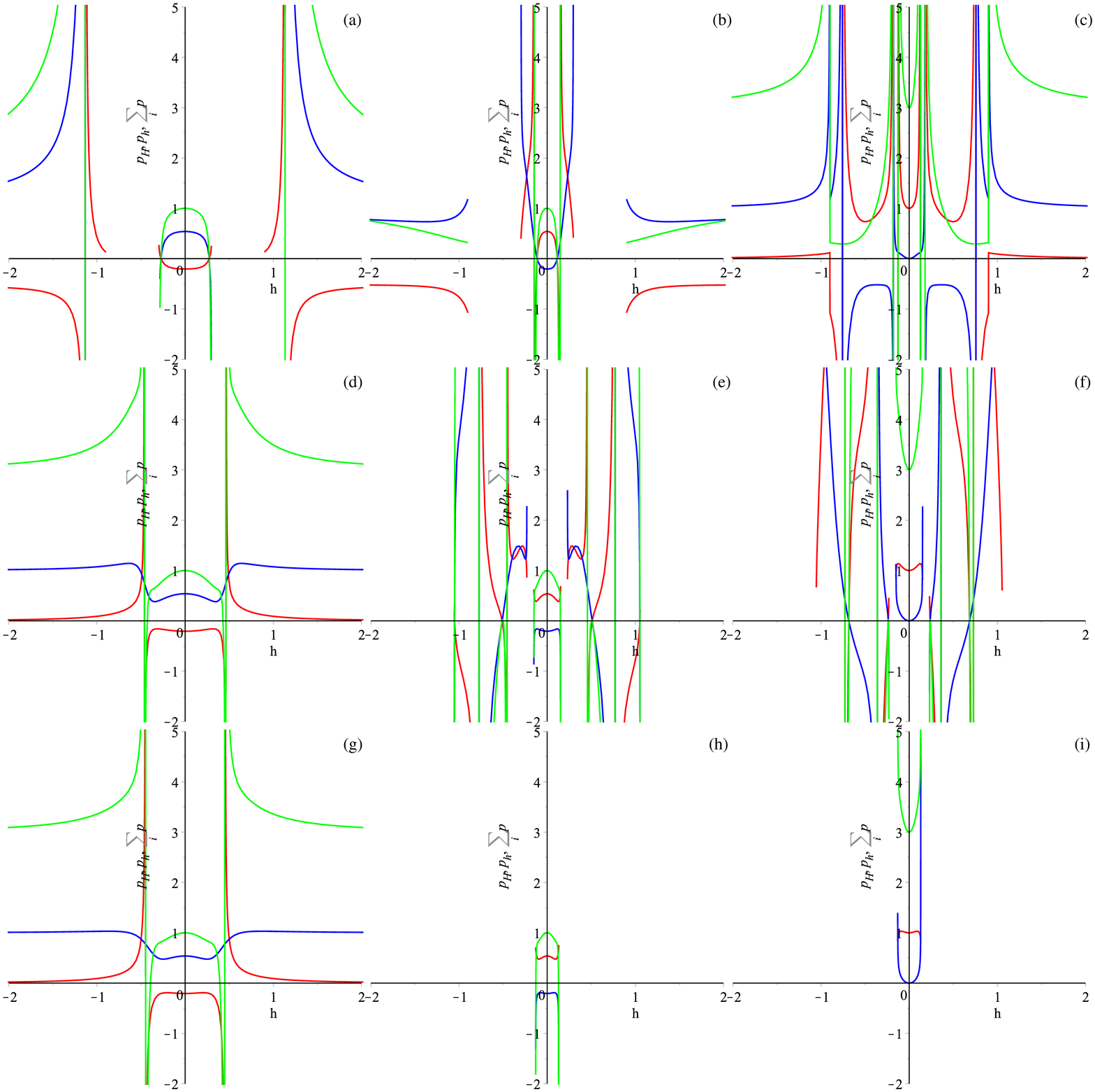}
\caption{$p_H$, $p_h$ and $\sum p_i$ graphs for vacuum $D=3$ case:  $p_H$ in red, $p_h$ in blue and $\sum p_i$ in green;
$\alpha > 0$, $\beta < 0$ on (a)--(c) panels: $H_1$ branch on (a) panel, $H_2$ branch on (b) panel and $H_3$ branch on (c) panel;
$\alpha > 0$, $\beta > 0$, $\mu < 0.3$ on (d)--(f) panels: $H_1$ branch on (d) panel, $H_2$ branch on (e) panel and $H_3$ branch on (f) panel;
$\alpha > 0$, $\beta > 0$, $\mu > 0.3$ on (g)--(i) panels: $H_1$ branch on (g) panel, $H_2$ branch on (h) panel and $H_3$ branch on (i) panel;
%$\alpha > 0$, $\beta < 0$ on (d) panel, $\alpha > 0$, $\beta > 0$, $\mu < 0.3$ on (e) panel, and $\alpha > 0$, $\beta > 0$, $\mu > 0.3$ on (f) panel
(see the text for more details).}\label{D3_ph_2}
\end{figure}

After that we obtain the expressions for the Kasner exponents, associated with both subspaces: $p_H = - H^2/\dot H$ and $p_h = - h^2/\dot h$. The $\dot H$ and $\dot h$ are obtained during the previous step while $H_i$ --
on pre-previous. Similar to the $\dot H$ and $\dot h$, the expressions themselves are bulky and we just provide the resulting graphs. They are presented in Figs.~\ref{D3_ph_1}--\ref{D3_ph_2} and the layout is following
that of $\dot H$ and $\dot h$: in Fig.~\ref{D3_ph_1} we
presented $\alpha < 0$ cases: $\beta < 0$ on (a)--(c) panels ($H_1$ branch on (a) panel, $H_2$ branch on (b) panel and $H_3$ branch on (c) panel),
$\beta > 0$, $\mu < 3/2$ on (d)--(f) panels ($H_1$ branch on (d) panel, $H_2$ branch on (e) panel and $H_3$ branch on (f) panel) and
$\beta > 0$, $\mu > 3/2$ on (g)--(i) panels ($H_1$ branch on (g) panel, $H_2$ branch on (h) panel and $H_3$ branch on (i) panel). In Fig.~\ref{D3_ph_2} we presented $\alpha > 0$ cases:
$\beta < 0$ on (a)--(c) panels ($H_1$ branch on (a) panel, $H_2$ branch on (b) panel and $H_3$ branch on (c) panel),
$\beta > 0$, $\mu < 0.3$ on (d)--(f) panels ($H_1$ branch on (d) panel, $H_2$ branch on (e) panel and $H_3$ branch on (f) panel) and
$\beta > 0$, $\mu > 0.3$ on (g)--(i) panels ($H_1$ branch on (g) panel, $H_2$ branch on (h) panel and $H_3$ branch on (i) panel).

\begin{figure}
\centering
\includegraphics[width=0.96\textwidth, angle=0]{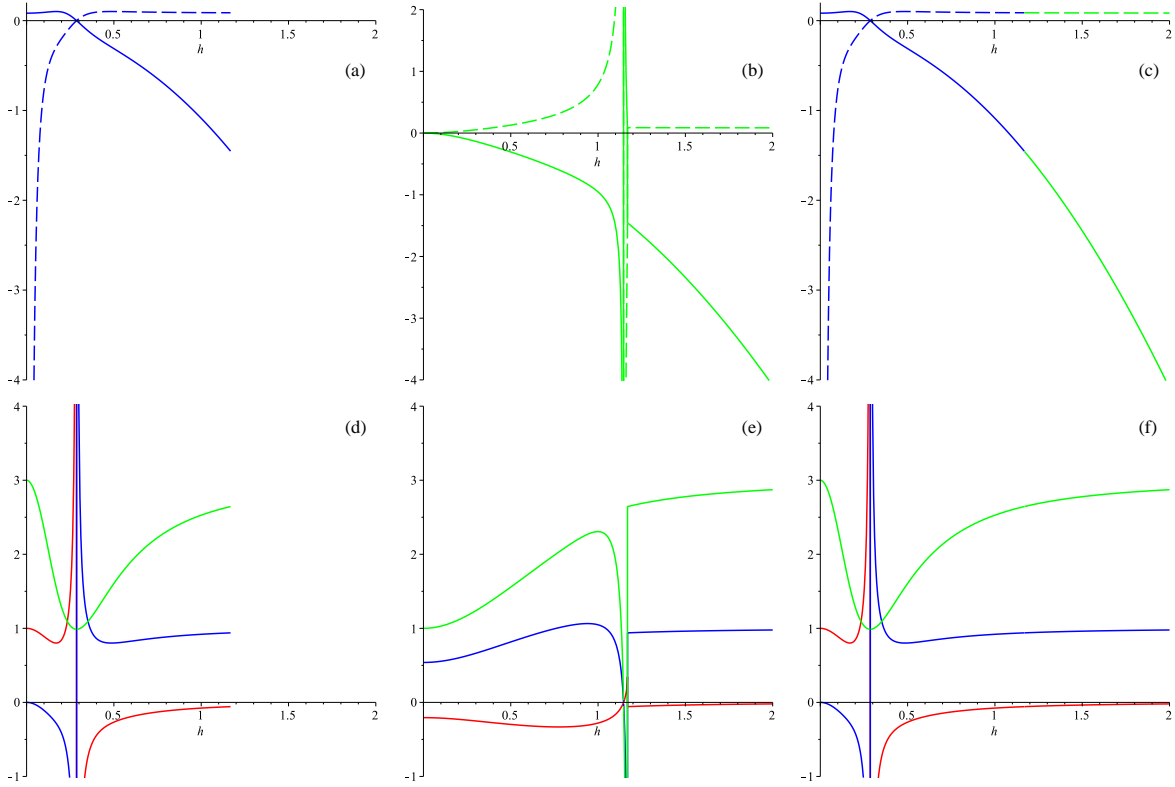}
\caption{The dynamics for $\alpha < 0$, $\beta < 0$ restored from gluing different branches: the case of $(\dot H(h), \dot h(h))$ on (a)--(c) panels and the case of Kasner exponents $(p_H, p_h)$ on (d)--(f) panels. On
(a) panel we presented $(\dot H(h), \dot h(h))$ for $H_2$ branch, on (b) -- for $H_3$ branch while in (c)---real physical evolution recovered from gluing proper parts of $H_2$ and $H_3$ branches. On (d)--(f) panels---the
same but for the Kasner exponents (red curve corresponds $p_H$, blue---to $p_h$ and green to $\sum p_i$)---coming from $H_2$ on (d), from $H_3$ on (e) and their combination on (f)
(see the text for more details).}\label{D3_2}
\end{figure}

Before proceeding with the description of the regimes, one more important note should be taken. As could be seen from Fig.~\ref{D3_1}, some of the branches are discontinued---for instance, let us consider $H_3$ branch from
Fig.~\ref{D3_1}(a) ($\alpha < 0$, $\beta < 0$ case). One can see that it describe ``internal'' part of the loop but starting from some $h > 0$ it ``jumps'' into another branch of evolution. Obviously, this cannot
happen in real physical evolution, so that it is the description which we use allows such ``jumps'', while the real physical evolution, for instance, for hyperbolic-like curve in the first quadrant of Fig.~\ref{D3_1}(a),
is described partially by $H_2$ and partially by $H_3$. Then to recover the real physical evolution, we ``glue'' the appropriate parts of the $(\dot H(h), \dot h(h))$ curves. As a way of example, for the above mentioned
hyperbolic-like curve from the first quadrant of Fig.~\ref{D3_1}(a), we plot $(\dot H(h), \dot h(h))$ for $H_2$ branch in Fig.~\ref{D3_2}(a) (see Fig.~\ref{D3_dHdh_1}(b)),
for $H_3$ branch---in Fig.~\ref{D3_2}(b) (see Fig.~\ref{D3_dHdh_1}(c)). Then we notice the discontinuity in
the Fig.~\ref{D3_2}(b)---it corresponds to the ``jump'' from the ``inner'' loop-like evolution curve to the curve under consideration at some $h=h_{cr}$. So that for $h < h_{cr}$ we use $H_2$ part and for $h > h_{cr}$---the
$H_3$ part; the resulting glued curve is presented in Fig.~\ref{D3_2}(c). One can verify that the resulting curve is free of any ``jumps'' and so represents physical evolution curve.

Exactly the same could be done for the analysis in terms of the Kasner exponents $p_i \equiv - H_i^2/\dot H_i$, and the example for the same curve is presented in Figs.~\ref{D3_2}(d)--(f). Again, in Fig.~\ref{D3_2}(d) we
presented Kasner exponents (red for $p_H$, blue for $p_h$ and green for $\sum p_i$) for $H_2$ branch (see Fig.~\ref{D3_ph_1}(b)), in Fig.~\ref{D3_2}(e)---for $H_3$ branch (see Fig.~\ref{D3_ph_1}(b))
(please note the discontinuity) and finally in Fig.~\ref{D3_2}(f) we present proper Kasner exponents for physical evolution curve.

With all these notes done, let us describe the resulting regimes. To do so we use $(\dot H(h), \dot h(h))$ curves from Figs.~\ref{D3_dHdh_1}--\ref{D3_dHdh_2} for all different cases, glue different branches properly to
obtain physical evolution curves and analyze them; the same procedure is performed for the Kasner exponents $p_H$ and $p_h$. We analyse the corresponding $(\dot H(h), \dot h(h))$ as well as $p_H$ and $p_h$ curves for
each particular case to find past and future asymptotes for all possible initial conditions.

 The resulting regimes are: for $\alpha < 0$,
$\beta < 0$ case (Figs.~\ref{D3_1}(a),~\ref{D3_dHdh_1}(a)--(c) and~\ref{D3_ph_1}(a)--(c)) we have $P_{1, 0} \toto E_{iso}$ and $P_{0, 1} \toto E_{iso}$ on hyperbolic-like curve and
$K_1 \toto nS$ as well as $nS \toto nS$ on the inner loop-like curve.
The regimes $P_{1, 0}$ and $P_{0, 1}$ are the regime with power-law behavior for one of the Hubble parameters with $p_i = 1$ and ``effective'' $p_j = 0$ for another, hence the designation: for $P_{1, 0}$ we have
$p_H = 1$, $p_h = 0$ while for $P_{0, 1}$ we have $p_H = 0$, $p_h = 1$. Let us note that for $D=3$ there is no real difference between $P_{1, 0}$ and $P_{0, 1}$ (since both subspaces are three-dimensional) but for
future cases there is, so we treat these two cases separately. Overall, in the $\alpha < 0$, $\beta < 0$ case there are no regimes with realistic compactifications.

The next case to consider is $\alpha < 0$, $\beta > 0$, $\mu < 3/2$ (Figs.~\ref{D3_1}(b),~\ref{D3_dHdh_1}(d)--(f) and~\ref{D3_ph_1}(d)--(f)), and there we have:
the ``outmost'' hyperbolic-like curve ($H_1$ branch) has $E_{iso}^1 \toto nS$, the ``middle'' hyperbolic-like curve
($H_2$ branch) has $P_{1, 0} \toto E_{iso}^2$ and $P_{0, 1} \toto E_{iso}^2$, while the ``innermost'' hyperbolic-like curve ($H_3$ branch) has $nS \toto K_1$. One can note that neither of the two considered so far cases has realistic regimes, as the only
non-singular regime has $P_{1, 0}, P_{0, 1} \toto E_{iso}$ and isotropic expansion of the entire space is not what we observe.

Let us move to the next case with $\alpha < 0$, $\beta > 0$, $\mu >  3/2$ (Figs.~\ref{D3_1}(c),~\ref{D3_dHdh_1}(g)--(i) and~\ref{D3_ph_1}(g)--(i)). One cannot miss the moment when $H_1$ and $H_2$ branches ``touch''
each other at $\mu = 3/2$; at that point two isotropic
exponential solutions $E_{iso}^{1, 2}$ coincide. For $\mu >  3/2$ the branches ``detach'' each other, forming banana-like curves as presented in Fig.~\ref{D3_1}(c). The isotropic exponential solutions degrade into
anisotropic ones, located on these banana-like curves. Similar to the previous case, the asymptotes for $H$ and $h$ are $\pm \sqrt{-\alpha/10\beta}$ and the corresponding regimes are nonstandard singularities. Then,
combining all above mentioned with the analysis of the $(\dot H(h), \dot h(h))$ curves allows us to conclude the regimes: $P_{0, 1} \toto E_{1}$, $P_{1, 0} \toto E_{2}$ and $nS \toto E_{1, 2}$ on banana-like curves and
$nS \toto K_1$ on central
curve. So that in this case we have $P_{1, 0}, P_{0, 1} \toto E_{1, 2}$ as a nonsingular regime, but $E_{1, 2}$ are located in the first and third quadrants, meaning either $(H > 0, h > 0)$ or $(H < 0, h < 0)$,
so that both three- and extra-dimensional spaces are expanding or contracting, which contradict our choice $(H > 0, h < 0)$, so that we discard this regime and cannot call it realistic.

We go on with $\alpha > 0$, $\beta < 0$ (Fig.~\ref{D3_1}(d),~\ref{D3_dHdh_2}(a)--(c) and~\ref{D3_ph_2}(a)--(c)) case. Similar to two previous cases, there are three limiting values for $H$ as
$h \to \pm\infty$ (as well as for $h$ as $H \to \pm\infty$) and they are the same
as for the previous case ($\pm \sqrt{-\alpha/10\beta}$), and, also similar to the previous cases, the corresponding regimes are nonstandard singularities. The anisotropic exponential solutions are located on edge-shaped
curves in second and fourth quadrants while isotropic---on hyperbola-like curve in first and third quadrants. Combining all these with the analysis of the $(\dot H(h), \dot h(h))$ curves allows us to draw the regimes:
$E_{2} \toto P_{1, 0}$, $E_{1} \toto P_{0, 1}$ and $E_{1, 2} \toto nS$ on the edge-shaped curves, $K_1 \toto nS$ and $nS \toto nS$ on central loop-like curve and $nS \toto E_{iso}$ on hyperbola-like curve.

In this case we have two interesting features which worth mentioning---first, we finally have realistic compactification---indeed, $P_{1, 0}, P_{0, 1} \to E_{1, 2}$ are realistic compactifications,
as both $E_{1, 2}$ have different
signs for $H$ and $h$\footnote{Actually, only one of them has $(H > 0, h < 0)$---another has $(H < 0, h > 0)$, but since it is $D=3$ case, for us it does not matter which of three-dimensional subspaces is expanding
and which is
contracting---we call the expanding one as ``our Universe'' and the contracting as extra dimensions, so that in $D=3$ instead of one we have two anisotropic exponential solutions.}. The second feature is inability to
reach $E_{iso}$ from the standard singularity---indeed, isotropic exponential solution is connected only to nonstandard singularity. This feature does not affect realistic compactification abundance, but we note it for
completeness; also, in the Gauss-Bonnet case~\cite{my16a, my16b, my17a, my18a} we never experienced such situation, neither in vacuum nor in $\Lambda$-term cases.

\begin{figure}
\centering
\includegraphics[width=0.96\textwidth, angle=0]{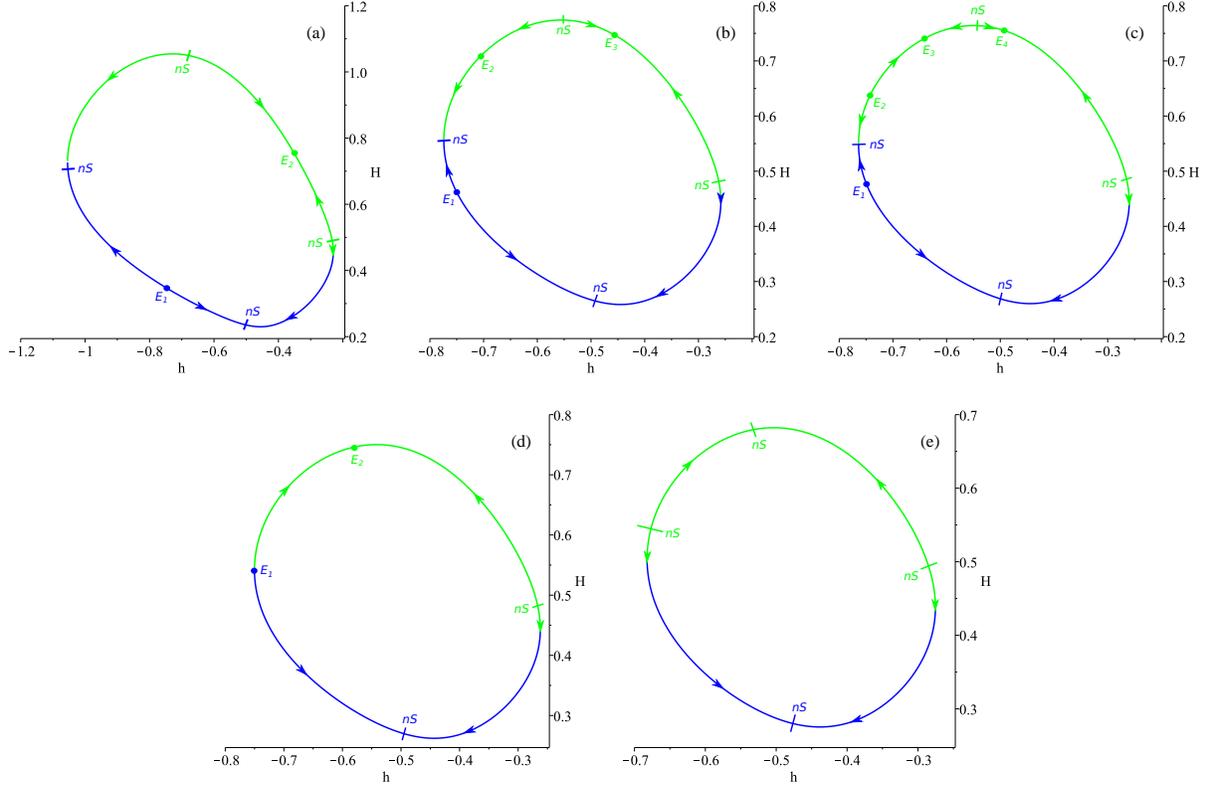}
\caption{The fine structure of the solutions in the $\alpha > 0$, $\beta > 0$ case: $\mu < 1/6$ on (a) panel, $\mu = 1/6$ on (b) panel, $\mu_3 > \mu > 1/6$ on (c) panel, $\mu = \mu_3$ on (d) panel and $0.3 > \mu > \mu_3$
on (e) panel. Different colors correspond to the different branches (blue -- to $H_2$ and green -- to $H_3$, in accordance with the designation in Fig.~\ref{D3_1})
(see the text for more details).}\label{D3_fine}
\end{figure}

The two remaining cases are similar---they both have $\alpha > 0$, $\beta > 0$ but one of them has $\mu < 0.3$ (Figs.~\ref{D3_1}(e),~\ref{D3_dHdh_2}(d)--(f) and~\ref{D3_ph_2}(d)--(f))
while another $\mu > 0.3$ (Figs.~\ref{D3_1}(f),~\ref{D3_dHdh_2}(g)--(i) and~\ref{D3_ph_2}(g)--(i)). The difference, as one can see from
Figs.~\ref{D3_1}(e, f) lies in circle-like curve in the second and fourth quadrants. The regime which is common for both cases is $P_{1, 0} \toto K_1$, and it is another example of the realistic compactification.
 The regimes within the circle-like curve in the second and fourth quadrants are subject to the ``fine-structure'' (see the description of the anisotropic exponential solutions
abundance above) and are presented in Fig.~\ref{D3_fine}. There we presented the following $\alpha > 0$, $\beta > 0$ cases: $\mu < 1/6$ on (a) panel, $\mu = 1/6$ on (b) panel, $\mu_3 > \mu > 1/6$ on (c) panel,
$\mu = \mu_3$ on (d) panel and $0.3 > \mu > \mu_3$ on (e) panel (at $\mu = 0.3$ the circle-like curve disappears). Different colors correspond to different branches, in accordance with the designation in Fig.~\ref{D3_1}) --
blue -- to $H_2$ and green -- to $H_3$. One can see that the abundance of the anisotropic exponential solutions exactly coinside with our description provided above.

\begin{figure}
\centering
\includegraphics[width=0.75\textwidth, angle=0]{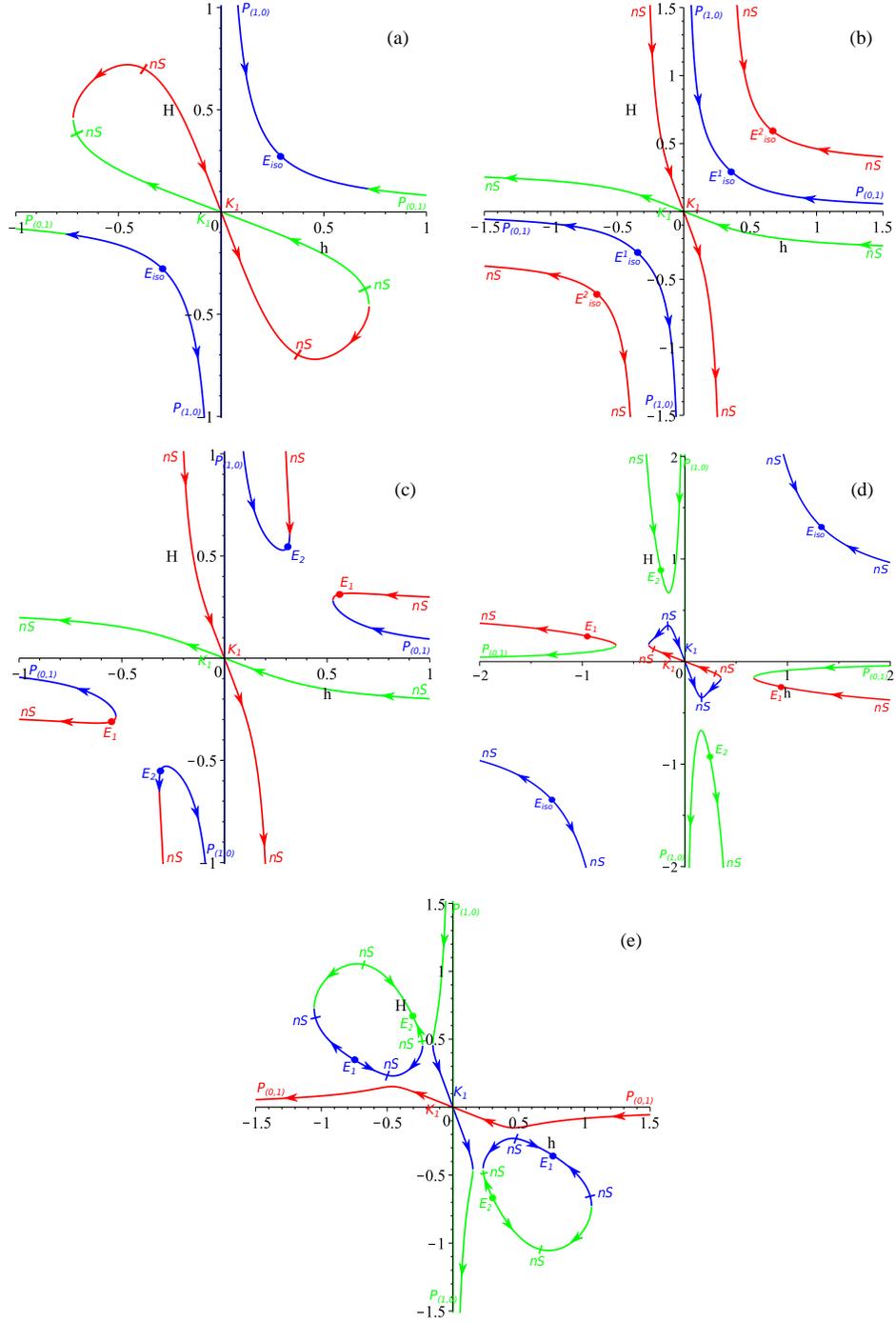}
\caption{Final compilations of all possible regimes in $D=3$ vacuum cubic Lovelock case, on $H(h)$ evolution curves; different colors correspond to three different branches $H_1$, $H_2$ and $H_3$;
panels layout is as follows:
$\alpha < 0$, $\beta < 0$ on (a) panel, $\alpha < 0$, $\beta > 0$, $\mu < 3/2$ on (b) panel, $\alpha < 0$, $\beta > 0$, $\mu > 3/2$ on (c) panel, $\alpha > 0$, $\beta < 0$ on
(d) panel, $\alpha > 0$, $\beta > 0$ and $\mu < 0.3$ on (e) panel
(see the text for more details).}\label{D3_all}
\end{figure}

For visualisation purposes we plot all discovered regimes on $H(h)$ curves, the results are presented in Fig.~\ref{D3_all}. The panels layout follow that of Fig.~\ref{D3_1}: $\alpha < 0$, $\beta < 0$ on (a) panel, $\alpha < 0$, $\beta > 0$, $\mu < 3/2$ on (b) panel, $\alpha < 0$, $\beta > 0$, $\mu > 3/2$ on (c) panel, $\alpha > 0$, $\beta < 0$ on
(d) panel, $\alpha > 0$, $\beta > 0$ and $\mu < 0.3$ on (e) panel; red curve corresponds to $H_1$, blue -- to $H_2$ and green -- to $H_3$. The arrows represent the evolution
according with respect to grow of the cosmic time $t$.

%\begin{figure}
%\includegraphics[width=1.0\textwidth, angle=0]{D3_all_1.eps}
%\caption{The dynamics for $\alpha < 0$, $\beta < 0$ restored from gluing different branches: the case of $(\dot H(h), \dot h(h))$ on (a)--(c) panels and the case of Kasner exponents $(p_H, p_h)$ on (d)--(f) branches. On
%(a) branch we presented $(\dot H(h), \dot h(h))$ for $H_2$ branch, on (b) -- for $H_3$ branch while in (c)---real physical evolution recovered from gluing proper parts of $H_2$ and $H_3$ branches. On (d)--(f) panels---the
%same but for the Kasner exponents (red curve corresponds $p_H$, blue---to $p_h$ and green to $\sum p_i$)---coming from $H_2$ on (d), from $H_3$ on (e) and their combination on (f)
%(see the text for more details).}\label{D3_all}
%\end{figure}

This concludes our consideration of the first and the simplest $D=3$ case. We concluded that there are two sets of realistic regimes: $P_{1, 0}, P_{0, 1} \to E_{1, 2}$ for $\alpha > 0$, $\beta < 0$
and $P_{1, 0}, P_{0, 1} \to K_1$ for $\alpha > 0$, $\beta > 0$;
interesting that both of them require $\alpha > 0$. Let us note that only half of these regimes have $H > 0, h < 0$ -- the other half have $H < 0, h > 0$, but since this particular case has $D=3$ -- the number of
extra dimensions coincide with the number of ``our'' spatial dimensions, we can ``switch'' between them, which effectively doubles the number of regimes.
One also can notice that our description based on the analysis of the discriminants of the equations provide accurate information on the regimes' abundance. Further, the
 analysis of the $H(h)$ and $(\dot H(h), \dot h(h))$ curves supplement this study with data on the non-standard singularities as well as findings, which of the exponential solutions cannot be reached and which cannot provide
 realistic compactification (having, for instance, both $H > 0$ and $h > 0$). So that on the particular example of $D=3$ case we demonstrated how our scheme works and are going to use it further without this much details.

\section{$D=4$ case}

In this case the equations of motion (\ref{H_gen})--(\ref{con2_gen}) take form ($H$-equation, $h$-equation, and constraint correspondingly)

\begin{equation}
\begin{array}{l}
4\dot H + 6H^2 + 8\dot h + 20h^2 + 16Hh + 8\alpha \( 2\dot H (4Hh + 6h^2) + 4\dot h (H^2 + 3h^2 + 6Hh) + \right.\\ \left. +8H^3h + 34H^2h^2  + 48Hh^3 + 15h^4 \) + 144\beta \( 2 (\dot H + H^2)(4Hh^3 + h^4) + \right.\\ \left.
+ 4(\dot h+ h^2)(3H^2h^2 + 2Hh^3)   + 5H^2h^4   \)
 = 0,
\end{array} \label{D4_H}
\end{equation}

\begin{equation}
\begin{array}{l}
6\dot H + 12H^2 + 6\dot h + 12h^2 + 18Hh + 8\alpha \( 3\dot H (H^2 + 6Hh + 3h^2) + 3\dot h (3H^2 +   6Hh + \right. \\ \left. + h^2)

+ 3H^4 + 27Hh^3  + 45H^2h^2 + 27H^3h + 3h^4 \) + 144\beta \( 3(\dot H + H^2) (3H^2h^2 + \right. \\ \left. + 2Hh^3)
+  3(\dot h + h^2)(2H^3h + 3H^2h^2) +5H^3h^3  \)
 = 0,
\end{array} \label{D4_h}
\end{equation}

\begin{equation}
\begin{array}{l}
6 H^2 + 24Hh + 12h^2 + 24\alpha (4H^3h + 18H^2h^2 + 12Hh^3 + h^4 ) + 720\beta ( 4H^3h^3 + 3H^2h^4) = 0.
\end{array} \label{D4_con}
\end{equation}

First we find the discriminant of (\ref{D4_con}) with respect to $H$ and then its discriminant with respect to $h$. The resulting equation is 15th order polynomial with respect to $\mu = \beta/\alpha^2$ and it has the
following roots: single roots

$$
\mu_1 = -4\sqrt[3]{98}/135 + 38/135 \approx 0.1449
$$

\noindent and $\mu_2 = 5/6$, as well as double root $0.3$ and triple root $\approx 0.4418$. We shall see that only first two roots, $\mu_1$ and $\mu_2$,
affect physical regimes.

To find isotropic exponential solutions, we substitute $\dot H = \dot h \equiv 0$ as well as $h=H$ into (\ref{D4_H})--(\ref{D4_con}), the system is simplified to a single equation

$$
%\begin{equation}
%\begin{array}{l}
42 H^2 \( 1 + 20\alpha H^2 + 120\beta H^4  \) = 0
%\end{array} \label{D4_iso1}
%\end{equation}
$$

\noindent with nontrivial solution

\begin{equation}
\begin{array}{l}
H^2 = \dac{-5\alpha \pm \sqrt{25\alpha^2 - 30\beta}}{60\beta}.
\end{array} \label{D4_iso2}
\end{equation}

Analyzing (\ref{D4_iso2}) we find out that there is one root if $\beta < 0$ (regardless of $\alpha$) and two roots if $\alpha < 0$, $5\alpha^2/6 \geqslant \beta > 0$; in all other cases there are no isotropic exponential solutions.

To find anisotropic exponential solutions abundance, we substitute $\dot H = \dot h \equiv 0$ into (\ref{D4_H})--(\ref{D4_con}) and solve the resulting system with respect to $H$ and $h$. The resulting equation on $h$ is
bi-eight-power and its discriminant is 16th order polynomial in $\mu$ with roots: single roots $\mu_3 = 841/5184 \approx 0.16223$, $\mu_4 = 1/6 \approx 0.16667$ and $\mu_5 = 5/6 \approx 0.8333$, as well as double roots
$\mu_{6,7} = 34/25 \pm 8\sqrt{14}/25 \approx 0.16267, 2.557$ and $\mu_{8,9} = 1/3 \pm \sqrt{21}/27 \approx 0.1636, 0.503$, as well as triple set of imagenary roots of fourth order equation. One can see that in the vicinity
of $\mu \approx 0.16\div 0.17$ there is a fine structure based on the rapid change of root number with change of $\mu$. Further analysis of the equations reveals the following: for $\alpha < 0$ and $\mu < 0$ we have
two roots for $\zeta = \alpha h^2$ and so for $h$ we have totally four roots (two positive and two negative) $\pm h_{1, 2}$; for $\alpha < 0$ and $\mu > 0$ there are no roots for $\mu < \mu_5 = 5/6$, for $\mu = 5/6$
there is a double root and for $\mu > 5/6$ there are two distinct roots $\pm h_{1, 2}$. So that for $\alpha < 0$ there are two stable exponential solutions for $\mu < 0$ and $\mu \geqslant 5/6$; for other $\mu$ there are
no anisotropic exponential solutions.

For $\alpha > 0$ the situation is more complicated: for $\mu < 0$, as in the $\alpha < 0$ case, there are two solutions, while for $\mu > 0$ it is much more complex. For $\mu < \mu_3$ there are four, $\mu = \mu_3$ five,
$\mu_8 > \mu > \mu_3$ six, $\mu = \mu_8$ again four, $\mu_4 > \mu > \mu_8$ two and for $\mu = \mu_4 = 1/6$ only one. For $\mu > 1/6$ there are no stable exponential solutions
for $\alpha > 0$. So that for $\alpha > 0$ there are anisotropic exponential solutions iff $\mu \leqslant 1/6$ (including $\mu < 0$ domain).

\begin{figure}
\centering
\includegraphics[width=0.7\textwidth, angle=0]{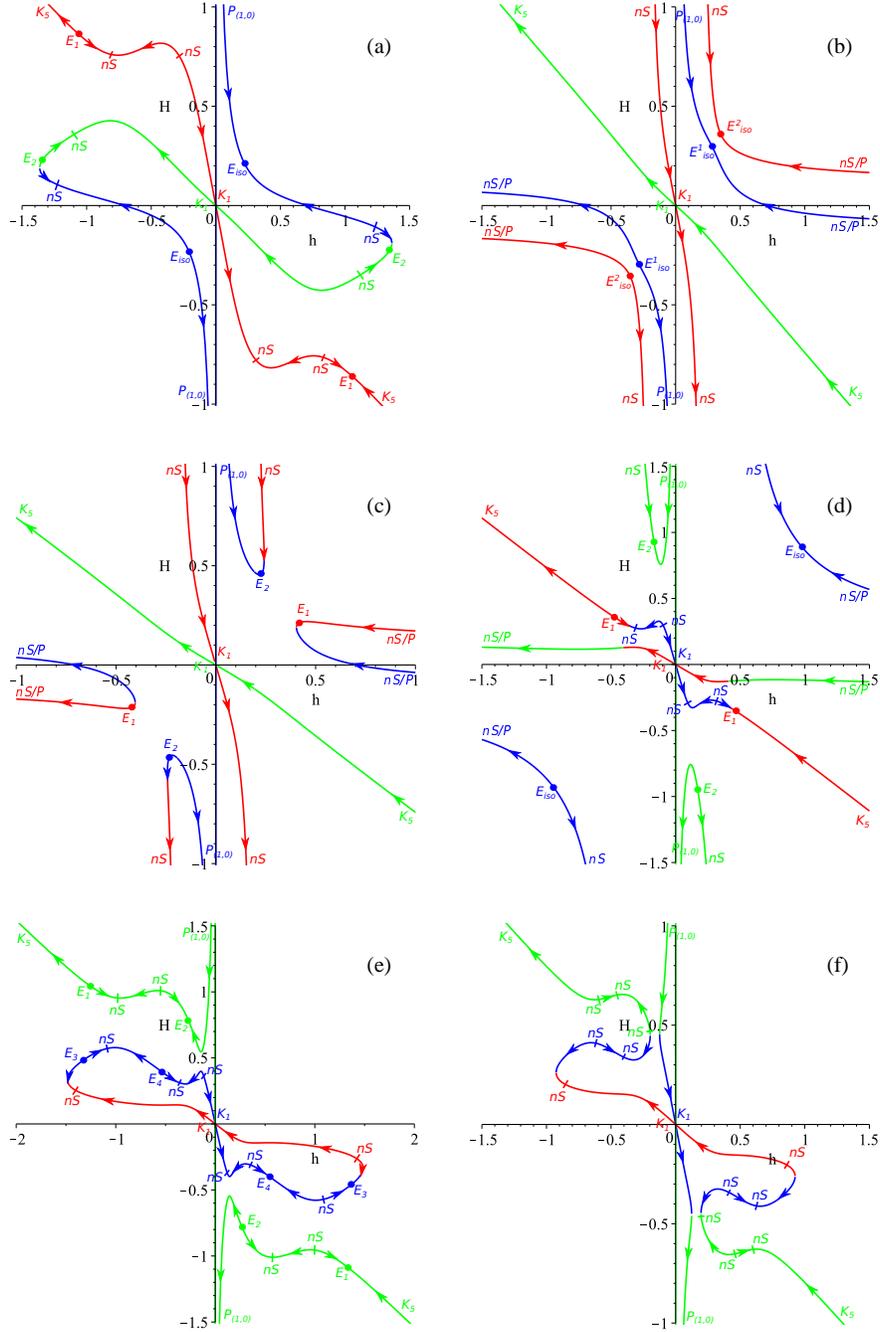}
\caption{Final compilations of all possible regimes in $D=4$ vacuum cubic Lovelock case, on $H(h)$ evolution curves; different colors correspond to three different branches $H_1$, $H_2$ and $H_3$;
panels layout is as follows: $\alpha < 0$, $\beta < 0$ on (a) panel, $\alpha < 0$, $\beta > 0$, $\mu < 5/6$ on (b) panel, $\alpha < 0$, $\beta > 0$, $\mu > 5/6$ on (c) panel, $\alpha > 0$, $\beta < 0$ on
(d) panel, $\alpha > 0$, $\beta > 0$, $\mu < \mu_1$ on (e) panel, and $\alpha > 0$, $\beta > 0$, $\mu > \mu_4$ on (f) panel
(see the text for more details).}\label{D4_1}
\end{figure}

Now it is time to explore $H(h)$ curves. They are presented in Fig.~\ref{D4_1}. There red curve corresponds to $H_1$, blue to $H_2$ and green
to $H_3$. The panels layout is as follows: $\alpha < 0$, $\beta < 0$ on (a) panel, $\alpha < 0$, $\beta > 0$, $\mu < 5/6$ on (b) panel, $\alpha < 0$, $\beta > 0$, $\mu > 5/6$ on (c) panel, $\alpha > 0$, $\beta < 0$ on
(d) panel, $\alpha > 0$, $\beta > 0$, $\mu < \mu_1$ on (e) panel, and $\alpha > 0$, $\beta > 0$, $\mu > \mu_4$ on (f) panel.
The situations on (b) and (c) panels coincide at $\mu = 5/6$ -- in that case the two isotropic exponential solutions coincide and the ``banana''-shaped areas ``touch'' each other at the point of double isotropic solution.
%The similar is the situation with (e) and (f) panels---the curves ``touch'' each other at $\mu=\mu_1$.

To proceed, we skip the intermediate steps (description of the individual $\dot H(h)$, $\dot h(h)$,
$p_H$, $p_h$ curves) and present the resulting regimes on the $H(h)$ curves (see Fig.~\ref{D4_1}).
Their analysis suggest the following: for $\alpha < 0$, $\mu < 0$ case (see Fig.~\ref{D4_1}(a)),
we have two nonsingular regimes: $K_5 \toto E_1$ regime on $H_1$ branch (red curve), and $P_{1, 0} \toto E_{iso}$ on $H_2$ (blue curve) as $h\to 0$ and $H\to\pm\infty$. One can see that in $D=4$ case, unlike $D=3$,
we have ``proper'' cubic-Lovelock Kasner
regime with $\sum p = 5$, as predicted. But as $E_1$ is located along $H_1$, it has either ($H < 0$, $h > 0$) or ($H > 0$, $h < 0$). Only one of them is of interest to us ($H > 0$, $h < 0$), but this regime is reachable
only as past asymptote: $E_1 \to K_5$; the $E_1$ from $K_5 \to E_1$ has ($H < 0$, $h > 0$). So that the regime which could give us realistic compactification, cannot be reached from $K_5$, despite it exists (but unstable). Singular regimes for $\alpha < 0$, $\mu < 0$ case include $E_{iso} \toto nS$, $E_2 \toto nS$, $nS \toto nS$ and $K_1 \toto nS$ along $H_2-H_3$ curve and $nS \toto E_1$, $nS \toto nS$ and $K_1 \toto nS$ along $H_1$ curve.
So that another anisotropic exponential solution $E_2$ is located between nonstandard singularities and cannot be
reached from the initial cosmological singularity.

Next case is $\alpha < 0$, $\beta > 0$, $\mu < 5/6$, which is presented in Fig.~\ref{D4_1}(b). There the only nonsingular regimes are $K_5 \toto K_1$ (along $H_3$, green line) and $P_{1, 0} \toto E_{iso}^1$.
Similar to $D=3$, $H_1$ and $H_2$
branches have fixed limit at
$h\to\pm\infty$, the corresponding regime is nonstandard singularity. This makes $E_{iso}^{2}$ to be connected only with $nS$; apart from these regimes there is also $K_1 \toto nS$. So that in this case we have
$K_5 \to K_1$ which is the transition from high-energy to low-energy Kasner, similar to the situation described in the Einstein-Gauss-Bonnet case~\cite{my16a}. But similar to the just described above the situation with
the anisotropic exponential solution, $K_1$ in this case has either ($H < 0$, $h > 0$) or ($H > 0$, $h < 0$). And again, similar to the previously described case, for $K_5 \to K_1$ we have ($H < 0$, $h > 0$) while for
$K_1 \to K_5$ we have ($H > 0$, $h < 0$). So that the Kasner regime which is suitable for us (three expanding and $D$ contracting dimensions) is unstable as $t\to\infty$. We shall discuss it more in the proper section
of the manuscript. Let us also note that $\mu = 5/6$ is the limiting case of (b) panel where the isotropic exponential solutions coincide.

The nonstandard singularities in this case have power-law behavior, that is why they are denoted as $nS/P$ -- they have singularity at $H = \pm \sqrt{-10\alpha/\beta}/30$ and they approach it in a power-law manner
in a finite time. So that the singularity remains nonstandard, yet, unlike other appearances, in this particular case its behavior is know, so that we specify it and denote as $nS/P$.

Now let us consider $\alpha < 0$, $\beta > 0$, $\mu > 5/6$ case, which is depicted in Fig.~\ref{D4_1}(c).  Similar to the previous case, as $h\to\pm\infty$ the values of $H$ tend to the same constants and the corresponding
regimes are again $nS/P$ for the same reasons as in the previous $\mu < 5/6$ case. Again, similar to the previous case, we have $K_5$ along $H_3$ (green line), but now isotropic exponential solutions ``turned'' into anisotropic, located on ``banana''-shaped curves.  So, similar to the
previous case, the we have $K_5 \toto K_1$ nonsingular regime and it is, again, have $H < 0$ for $K_1$ so it is invalid for our purpose of finding proper compactification. But unlike the previous case, we also have
$P_{1, 0} \toto E_2$ -- transition to the anisotropic exponential solution---another nonsingular regime. But this anisotropic solution is located in the first and third quadrants and so it has either ($H > 0$, $h > 0$)
or ($H < 0$, $h < 0$) -- so both three
and extra dimensions are either expanding or contracting and so we cannot explain compactification with this regime. The singular regimes for this case include $E_1 \toto nS/P$, $E_2 \toto nS$ and $K_1 \toto nS$.

So that in these three cases which combine entire $\alpha < 0$ domain there is no realistic compactification. Let us continue with $\alpha > 0$ regimes. First one to consider is $\alpha > 0$, $\beta < 0$, presented in
Fig.~\ref{D4_1}(d). There we have isotropic exponential solution on $H_2$ (blue hyperbola-like curve) and $K_5$ on $H_1$ (red line), but unlike previous cases there is no $K_5 \toto K_1$ transition, as there is
anisotropic exponential solution on $H_1$ as well, so we have $K_5 \toto E_1$ instead. And again, similar to one of the previous cases, this exponential solution has $H < 0$ on $K_5 \to E_1$ and $H > 0$ on $E_1 \to K_5$,
so that we cannot
describe realistic compactification with this regime.
Similar to $D=3$ case the isotropic exponential solution is ``surrounded'' by nonstandard singularities -- the regimes involve $E_{iso}$ are $E_{iso} \toto nS$ and $E_{iso} \toto nS/P$.
Another non-singular regime is $P_{1, 0} \toto E_2$ which takes place on $H_3$ ``edge''-shaped part, and in this case $E_2$ has $H > 0, h < 0$ and is stable past-time
asymptote, so that it could describe the compactification.
The singular regimes in this case include $E_1 \toto nS$,
$E_2 \toto nS$, $K_1 \toto nS/P$ and $K_1 \toto nS$. To conclude, this case provides us with $P_{1, 0} \to E_2$ which could describe realistic compactification.

\begin{figure}
\centering
\includegraphics[width=0.7\textwidth, angle=0]{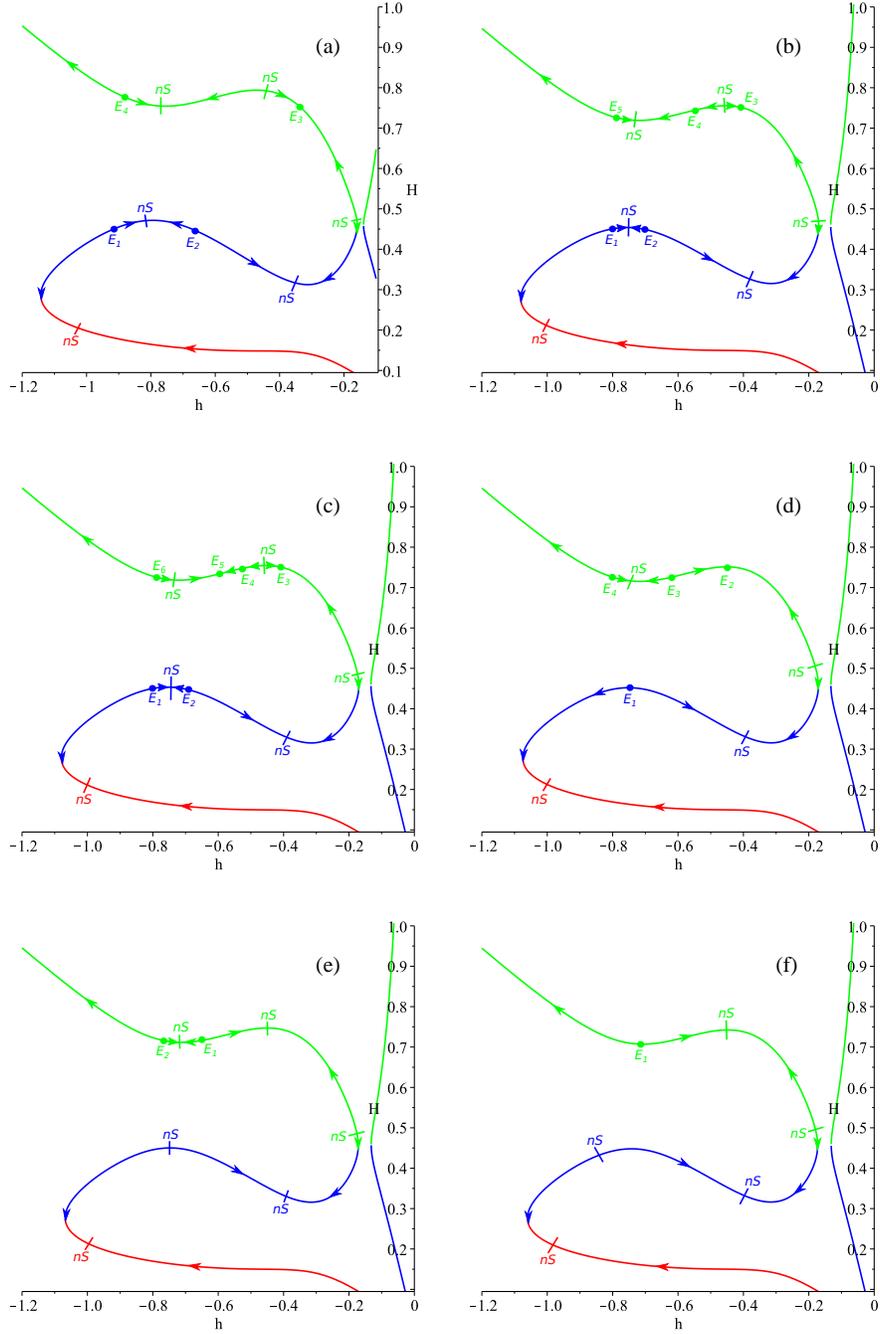}
\caption{The fine structure of the solutions in the $\alpha > 0$, $\beta > 0$ $D=4$ vacuum case: $\mu_1 < \mu < \mu_3$ on (a) panel, $\mu = \mu_3$ on (b) panel, $\mu_8 > \mu > \mu_3$ on (c) panel, $\mu = \mu_8$ on (d) panel,
$1/6 > \mu > \mu_8$
on (e) panel and $\mu = \mu_4 = 1/6$ on (f) panel.
Different colors correspond to the different branches (red -- to $H_1$, blue -- to $H_2$ and green -- to $H_3$, in accordance with the designation in Fig.~\ref{D4_1})
(see the text for more details).}\label{D4_2}
\end{figure}

The last case to consider is $\alpha > 0$, $\beta > 0$, and it has wide variety of the regimes, as well as fine structure, similar to the previous $D=3$ case. But unlike it, now all the regimes from the fine
structure are connected with the initial singularity, making them more physical. We presented two representative cases -- $\mu < \mu_1$ and $\mu > \mu_4 = 1/6$ in Figs~\ref{D4_1}(e) and~\ref{D4_1}(f) respectively, while all
cases inbetween -- in Fig.~\ref{D4_2}.

Let us start the description with growth of $\mu$. The first case is $\mu < \mu_1$, presented in Fig.~\ref{D4_1}(e). The regimes along $H_3$ (green curve) in the second quadrant are: \linebreak
\mbox{$K_5 \ot E_1 \to nS \ot nS \to E_2 \ot P_{1, 0}$}; the regimes in the fourth quadrant are time-reverse of the described. So that $E_1$ there is past asymptote and cannot describe compactification, while $E_2$ case, and
$P_{1, 0} \to E_2$ is an example of the compactification regime. The regimes along eight-like figure formed from $H_1$ and $H_2$ branches in the center are (starting from red $H_1$ branch):
$K_1 \to nS \ot E_3 \to nS \ot E_4 \to nS \ot nS \to K_1$, where first $K_1$ belong to $H_1$ branch while the last -- to $H_2$. One can see that there are two anisotropic exponential solutions along this curve, but
both of them are unstable. So that in this case only $P_{1, 0} \to E_2$ is compactification regime. The next case is $\mu = \mu_1$ -- in that case $H_2$ (blue curve) and $H_3$ (green curve) ``touch'' each other, but the
regimes remain unchanged. With further increase of $\mu$ the $H_2$ and $H_3$ branches ``detouch'' each other but in a way presented in Fig.~\ref{D4_2}(a), where we presented fine-structure (and so only the second
quadrant) for $\mu_3 > \mu > \mu_1$. One can see that now $H_3$ is ``cut'' into two physical branches -- one of them together with a piece from $H_2$ is forming $P_{1, 0} \to K_1$ regime (see also Fig.~\ref{D4_1}(f) where
it seen clearly). Another part of $H_3$ is forming with remaining parts of $H_2$ and $H_1$ the second physical branch with the regimes (starting from $K_1$ on $H_1$):
$K_1 \to nS \ot E_1 \to nS \ot E_2 \to nS \ot nS \to E_3 \ot nS \to nS \ot E_4 \to K_5$. One can see there four different anisotropic exponential solution (similar as in $D = 3$ case, see~Fig.~\ref{D3_fine}(c))
but now located along the same physical
branch. Three out of them are unstable ($E_1$, $E_2$ and $E_4$) and the remaining stable $E_3$ is ``surrounded'' by $nS$ and cannot be a part of compactification scenario. To conclude, the only compactification regime here is
$P_{1, 0} \to K_1$.

The further growth of $\mu$ to $\mu = \mu_3$ is presented in Fig.~\ref{D4_2}(b) -- the situation and the regimes are almost the same with the difference that now an additional exponential solution ($E_4$ by designations
on Fig.~\ref{D4_2}(b)) emerged ``between'' two nonstandard singularities on the upper part of $H_3$. Since it emerged between $nS$, no new physically significant regimes are formed. With increase of $\mu$ this exponential
solution, which has $\sum H_i = 0$ (constant-volume exponential solution, see~\cite{CST2}) at $\mu=\mu_3$, is split into two for $\mu_8 > \mu > \mu_3$ -- one of them ($E_2$) is stable while another ($E_3$) is unstable
(see  Fig.~\ref{D4_2}(c) for illustration). But again, both of them are surrounded by nonstandard singularities so no new physically sensitive regimes appear.

The next qualitative change of the situation happening at $\mu = \mu_8$, displayed in Fig.~\ref{D4_2}(d). At this point two changes happening: at $H_2$ (blue segment) two exponential solutions emerged with nonstandard
singularity forming single exponential solution $E_1$; similar situation happened at $H_3$ (green segment). But still, all these changes are within $nS$ bounds, so, similar to the previous cases, no changes of the
realistic regimes occur. With further growth of $\mu$ to $\mu_4 > \mu > \mu_8$ these newly formed exponential solutions turn to nonstandard singularities (see Fig.~\ref{D4_2}(e)); the remaining two exponential solutions
($E_1$ and $E_2$ on the $H_3$) merge with nonstandard singularity between them into new exponential solution at $\mu=\mu_4$ (see Fig.~\ref{D4_2}(f)). Finally, at $\mu > \mu_4$ (Fig.~\ref{D4_1}(f)) no exponential solution
remains and there are only $nS \toto nS$ and $K_5 \toto nS$ along $H_2-H_3$ physical branch. So that for the entire $\mu > \mu_1$ we have only $P_{1, 0} \to K_1$ as realistic compactification regime; all the regimes within
fine structure are non-realistic.

To conclude, in $D=4$ we encounter proper cubic Lovelock Kasner regime $K_5$ with $\sum p = 5$, but there are no realistic compactification regimes originating from $K_5$. Instead,  there are $P_{1, 0} \to E_{3+4}$ for
$\alpha > 0$, $\mu \leqslant \mu_1$ and $P_{1, 0} \to K_1$ for $\alpha > 0$, $\mu > \mu_1$. So that for entire $\alpha > 0$ we have either of these regimes, and, similar to the previous cases, have no realistic regimes
for $\alpha < 0$.

\section{Discussions}

In the current paper we have analyzed the cosmological dynamics of the cubic Lovelock gravity, with Einstein-Hilbert and Gauss-Bonnet terms present as well. We have chosen a setup with a topology being a product of two
isotropic subspaces -- three-dimensions, representing our Universe, and $D$-dimensional, representing extra dimensions. Both subspaces are flat, which simplifies our equations of motion and makes it possible to analyze them
analytically. In a sense, it is a logical continuation of~\cite{my16a, my16b, my17a, my18a}, where we considered the same problem but in EGB gravity -- vacuum case in~\cite{my16a, my18a} and $\Lambda$-term case in~\cite{my16b, my17a, my18a}. In~\cite{my18a} we reviewed all the results for EGB from~\cite{my16a, my16b, my17a} and changed the visualization of the regimes -- in the original papers~\cite{my16a, my16b, my17a} we use tables to list of all the regimes, and this way sometimes is not easy to read. On the contrary, in~\cite{my18a} we put all
the regimes on $H(h)$ curves and added arrows to demonstrate $t\to\infty$ directed evolution. In the current paper we decided to keep visualization from~\cite{my18a}.

First of all, let us summarize the results, as they are scattered over mini-conclusions in each particular section. The fist one is $D=3$ and it has interesting feature -- since the
equations of motion are cubic in both $H$ and $h$, there could be up to three branches of the solutions. On the other hand, it is the lowest possible dimension for cubic Lovelock
gravity, so there are no Kasner solutions (see~\cite{PT}). Then the only possibility is what we call ``generalized Taub'' solution -- the situation when the expansion in each direction
is characterized by Kasner exponent $p_i = -H_i^2/\dot H_i$ equal to either 1 or 0; so that for our topology it is either $P_{1, 0}$ ($p_H = 1, p_h = 0$ -- expansion of the three-dimensional subspace and ``static'' extra dimensions) or $P_{0, 1}$ ($p_H = 0, p_h = 1$ -- expansion of the extra-dimensional subspace and ``static'' three dimensions). Then
the remaining branches -- which cannot be connected to either $P_{1, 0}$ or $P_{0, 1}$, form closed evolution curves for ($\alpha < 0, \beta < 0$) (see Fig.~\ref{D3_all}(a)) or
($\alpha > 0, \beta > 0$) (see Figs.~\ref{D3_all}(e) and~\ref{D3_fine}); for ($\alpha < 0, \beta > 0$) and ($\alpha > 0, \beta < 0$) they encounter nonstandard singularities (see Figs.~\ref{D3_all}(b)--(d)).
The realistic compactification regimes are $P_{1, 0}/P_{0, 1} \to E_{3+3}$ for ($\alpha > 0, \beta < 0$) and
 $P_{1, 0}/P_{0, 1} \to K_1$ for ($\alpha > 0, \beta > 0$); let us note that both of the regimes exist only for $\alpha > 0$.

The $D=4$ case has one cubic Lovelock Kasner solution $K_5$ but it is still not enough for all branches, so we still nonstandard singularities at ($\alpha < 0, \beta > 0$) and
($\alpha > 0, \beta < 0$) (see Figs.~\ref{D4_1}(b)--(d)) while for ($\alpha < 0, \beta < 0$) and ($\alpha > 0, \beta > 0$) the evolution curves have complicated shapes
(see Figs.~\ref{D4_1}(a), (e)--(f) and~\ref{D4_2}). In $D=4$ we still have $P_{1, 0}$ regime, but not $P_{0, 1}$, and some of the nonstandard singularities have power-law behavior
and so designated as $nS/P$. Unlike $D=3$, where the regimes within the fine structure existed on an isolated $H(h)$ curve (see Fig.~\ref{D3_fine}), in $D=4$ they are located on
one of the physical branches connected with $K_1$.
The realistic compactification regimes are $P_{1, 0} \to E_{3+4}$ for $\alpha > 0$, $\mu < \mu_1$ (including entire $\beta < 0$) and $P_{1, 0} \to K_1$, for $\alpha > 0$, $\mu > \mu_1$ -- exactly the same as in $D=3$ case, and again both of the regimes exist only for $\alpha > 0$.

The above-mentioned ``generalized Taub'' solution deserves some additional comments. Formally it fits ``generalized Milne'' solution -- the second branch of the power-law solutions
in Lovelock gravity (see~\cite{prd09} for details), but it is only formal -- it fits only because it is degenerative. As it was demonstrated in~\cite{PT}, strict ``generalized Milne''
cannot exist in pure highest-order Lovelock gravity, as it leads to degeneracy in the equations of motion. But if additional (lower-order) Lovelock contributions are involved, it could
restore this branch of solutions, but it was never demonstrated exactly before. So that on the particular example of $[3+D]$ spatial splitting we demonstrated this possibility. Still,
a little is known about this regime and it deserves additional investigation in the separate papers.

When we consider this ``generalized Taub'' solution as a past asymptote -- and this is the case for all possible realistic compactification models in $D=3, 4$ -- it feels a bit
unnatural. Indeed -- the $P_{1, 0}$ regime imply $H \to \infty$ and $h \to 0$ as $t \to 0$ (by ``0'' we mean here initial cosmological singularity), so that we initially have ``burst''-like
expansion of three-dimensional subspace while the extra-dimensional subspace is almost static. In addition to the feeling of unnaturalness, it is a question if this regime could be
reached from totally anisotropic space, in a manner it was done in~\cite{PT2017}  for EGB case. So that it gives additional reason to deeply investigate this regime and we are going
to do it in the nearest time.

The results of our analysis suggest that the variety and abundance of the regimes is closer to $\Lambda$-term EGB, rather then vacuum EGB models. The reasons for that are not clear, but we expect that number of the free
parameters plays a role here. Indeed, for vacuum EGB model there is only one parameter -- $\alpha$, Gauss-Bonnet coupling, while for $\Lambda$-term EGB and vacuum cubic Lovelock there are two -- $\alpha$ and $\Lambda$ for
the former and $\alpha$ and $\beta$ (cubic Lovelock coupling) for the latter. In that case the dynamics of the $\Lambda$-term cubic Lovelock gravity would be even more interesting and
we are going to consider this case shortly.

\section{Conclusions}

This concludes our study of the cosmological models in vacuum cubic Lovelock gravity with $D=3, 4$ extra dimensions.
We have found that in both cases there are regimes with successful compactification, but
all of them originate from ``generalized Taub'' solution; for the future asymptote we have either Kasner regime or anisotropic exponential solution.

Apart from the regimes with successful compactification, we described and plotted on $H(h)$ curves all possible regimes for all initial conditions and all structurally different cases.
The variety and abundance of the regimes exceed even $\Lambda$-term EGB case, featuring transition between two anisotropic exponential solutions and transition between two different
``generalized Taub'' solutions.

There are two interesting observations which require additional investigation, as both are quite unexpected. First of them
 is that all of the regimes with realistic compactification have $\alpha > 0$ requirement. This is totally unexpected, as in both vacuum and $\Lambda$-term EGB cases we have viable
 compactifications for both signs of $\alpha$. For the $\Lambda$-term case the joint analysis of our cosmological bounds and those coming from AdS/CFT and other considerations allows
 us to conclude $\alpha > 0$ (see~\cite{my16b, my18a}), but for that we involved external (to our results) analysis. In this case without any external bounds we already have
 realistic compactification only for $\alpha > 0$.

 The second observation is that there is no $K_5 \to K_1$ transition with realistic compactification.  In EGB vacuum case~\cite{my16a, my18a} we have the transitions of this kind, so it was natural to assume that in higher-order Lovelock gravity they also present, but our investigation reveals that they are not. There is $K_5 \to K_1$ transition, but with contracting
 three and expanding extra dimensions, so it formally exist, but with no compactification scenario. As both of these observations are unexpected and in disagreement with what we have
 learned from study of EGB case, this is a good direction for further improvement of our understanding of Lovelock gravity.


\begin{thebibliography}{99}
% Reference 1



\bibitem{Nord1914} G. Nordstr\"om, Phys. Z. {\bf 15}, 504 (1914).

\bibitem{Nord_2grav} G. Nordstr\"om, Ann. Phys. (Berlin) {\bf 347}, 533 (1913).

\bibitem{einst} A. Einstein, Ann. Phys. (Berlin) {\bf 354}, 769 (1916).

\bibitem{KK1} T. Kaluza, Sit. Preuss. Akad. Wiss. {\bf K1}, 966 (1921).

\bibitem{KK2} O. Klein, Z. Phys. {\bf 37}, 895 (1926).

\bibitem{KK3} O. Klein, Nature (London) {\bf 118}, 516 (1926).

\bibitem{sch-sch} J. Scherk and J.H. Schwarz, Nucl. Phys. {\bf B81}, 118 (1974).

\bibitem{VSh1} M.A. Virasoro, Phys. Rev. {\bf 177}, 2309 (1969).

\bibitem{VSh2} J.A. Shapiro, Phys. Lett. {\bf 33B}, 361 (1970).

\bibitem{Candelas_etal} P. Candelas, G.T. Horowitz, A. Strominger and E. Witten, Nucl. Phys. {\bf B258}, 46 (1985).

\bibitem{Gross_etal} D.J. Gross, J. Harvey, E. Martinec and R. Rohm, Phys. Rev. Lett. {\bf 54}, 502 (1985).

\bibitem{zwiebach} B. Zwiebach, Phys. Lett. {\bf 156B}, 315 (1985).

\bibitem{Lanczos1} C. Lanczos, Z. Phys. {\bf 73}, 147 (1932).

\bibitem{Lanczos2} C. Lanczos, Ann. Math. {\bf 39}, 842 (1938).

\bibitem{zumino} B. Zumino, Phys. Rep. {\bf 137}, 109 (1986).

\bibitem{Lovelock} D.~Lovelock, J. Math. Phys. (N.Y.) \textbf{12}, 498 (1971).


\bibitem{add_1} F. M${\ddot {\rm u}}$ller-Hoissen, Phys. Lett. {\bf 163B}, 106 (1985).

\bibitem{Deruelle2} N. Deruelle and L. Fari\~na-Busto, Phys. Rev. D {\bf 41}, 3696 (1990).

\bibitem{add_4} F. M${\ddot {\rm u}}$ller-Hoissen, Class. Quant. Grav. {\bf 3}, 665 (1986).











\bibitem{prd09} S.A. Pavluchenko,   Phys. Rev. D {\bf 80}, 107501 (2009).


\bibitem{add_10} J. Demaret, H. Caprasse, A. Moussiaux, P. Tombal, and D. Papadopoulos,  Phys. Rev. D {\bf 41}, 1163 (1990).

\bibitem{add_8} G. A. Mena Marug\'an, Phys. Rev. D {\bf 46}, 4340 (1992).

\bibitem{add13} E. Elizalde, A.N. Makarenko, V.V. Obukhov, K.E. Osetrin, and A.E. Filippov,  Phys. Lett. {\bf B644}, 1 (2007).

\bibitem{MO04} K.I. Maeda and N. Ohta, Phys. Rev. D \textbf{71}, 063520 (2005).

\bibitem{MO14} K.I.~Maeda and N.~Ohta, JHEP {\bf 1406}, 095 (2014).

\bibitem{add_rec_1} J.T. Wheeler, Nucl. Phys. {\bf B268}, 737 (1986).

%\bibitem{add_rec_2} R.G. Cai, Phys. Rev. D {\bf 65}, 084014 (2002).

\bibitem{addn_1} T. Torii and H. Maeda, Phys. Rev. D {\bf 71}, 124002 (2005).

\bibitem{addn_2} T. Torii and H. Maeda, Phys. Rev. D {\bf 72}, 064007 (2005).

\bibitem{add_rec_2} R.G. Cai, Phys. Rev. D {\bf 65}, 084014 (2002).

\bibitem{alpha_12}
D.G. Boulware and S. Deser,
%{\it String-generated gravity models},
Phys. Rev. Lett. {\bf 55}, 2656 (1985).



\bibitem{alpha_12}
Boulware, D.G. and Deser, S. String-generated gravity models.
%{\it String-generated gravity models},
{\em Phys. Rev. Lett.} {\bf 1985}, {\em 55}, 2656--2660.


\bibitem{add_rec_3} D.L. Wilshire. Phys. Lett. {\bf B169}, 36 (1986).

\bibitem{add_rec_4} R.G. Cai, Phys. Lett. {\bf 582}, 237 (2004).



\bibitem{addn_3} J. Grain, A. Barrau, and P. Kanti, Phys. Rev. D {\bf 72}, 104016 (2005).

\bibitem{addn_4} R. Cai and N. Ohta, Phys. Rev. D {\bf 74}, 064001 (2006).

\bibitem{addn_4.1} X.O. Camanho and J.D. Edelstein, Class. Quant. Grav. {\bf 30}, 035009 (2013).


\bibitem{addn_5} H. Maeda, Phys. Rev. D {\bf 73}, 104004 (2006).

\bibitem{addn_6} M. Nozawa and H. Maeda, Class. Quant. Grav. {\bf 23}, 1779 (2006).

\bibitem{addn_7} H. Maeda, Class. Quant. Grav. {\bf 23}, 2155 (2006).

\bibitem{addn_8} M. Dehghani and N. Farhangkhah, Phys. Rev. D {\bf 78}, 064015 (2008).

\bibitem{Is86} H. Ishihara, Phys. Lett. \textbf{B179}, 217 (1986).



\bibitem{Deruelle1} N. Deruelle, Nucl. Phys. {\bf B327}, 253 (1989).

\bibitem{mpla09} S.A. Pavluchenko and A.V. Toporensky, Mod. Phys. Lett. {\bf A24}, 513 (2009).



\bibitem{prd10} S.A. Pavluchenko,   Phys. Rev. D {\bf 82}, 104021 (2010).

\bibitem{Ivashchuk} V. Ivashchuk, Int. J. Geom. Meth. Mod. Phys. {\bf 07}, 797 (2010) \href{http://arxiv.org/abs/0910.3426v3}{arXiv: 0910.3426}.

\bibitem{grg10}   I.V. Kirnos, A.N. Makarenko, S.A. Pavluchenko, and A.V. Toporensky, General Relativity and Gravitation {\bf 42}, 2633 (2010).

\bibitem{PT} S.A. Pavluchenko and A.V. Toporensky, Gravitation and Cosmology {\bf 20}, 127 (2014); \href{http://arxiv.org/abs/1212.1386}{arXiv:
    1212.1386}.

\bibitem{KPT} I.V. Kirnos, S.A. Pavluchenko, and A.V. Toporensky, Gravitation and Cosmology {\bf 16}, 274 (2010)
    \href{http://arxiv.org/abs/1002.4488v2}{arXiv: 1002.4488}.

\bibitem{CPT1} D. Chirkov, S. Pavluchenko, A. Toporensky, Mod. Phys. Lett. {\bf A29}, 1450093 (2014); \href{http://arxiv.org/abs/1401.2962}{arXiv:
    1401.2962}.

\bibitem{CST2} D. Chirkov, S. Pavluchenko, A. Toporensky, Gen. Rel. Grav. {\bf 46}, 1799 (2014); \href{http://arxiv.org/abs/1403.4625}{arXiv: 1403.4625}.

\bibitem{CPT3} D. Chirkov, S. Pavluchenko, A. Toporensky, Gen. Rel. Grav. {\bf 47}, 137 (2015); \href{http://arxiv.org/abs/1501.04360}{arXiv:1501.04360}.

\bibitem{my15} S.A. Pavluchenko,   Phys. Rev. D {\bf 92}, 104017 (2015).

\bibitem{iv16} V. D. Ivashchuk, Eur. Phys. J. C {\bf 76}, 431 (2016).

%\bibitem{iv18} Ivashchuk, V.D.; Kobtsev, A.A. Stable exponential cosmological solutions with $3$- and $l$-dimensional factor spaces in the Einstein-Gauss-Bonnet model with a $\Lambda$-term.
%{\em Eur. Phys. J. C} {\bf 2018}, {\em 78}, 100.

\bibitem{CGP1} F.~Canfora, A.~Giacomini and S.~A.~Pavluchenko,  Phys.\ Rev.\ D {\bf 88}, 064044 (2013).

\bibitem{CGP2} F.~Canfora, A.~Giacomini and S.~A.~Pavluchenko,  Gen. Rel. Grav. {\bf 46}, 1805 (2014).

\bibitem{CGPT} F.~Canfora, A.~Giacomini, S.~A.~Pavluchenko and A. Toporensky,
%{\it Friedmann dynamics recovered from compactified Einstein-Gauss-Bonnet cosmology},
%\href{http://arxiv.org/abs/1605.00041}{arXiv:1605.00041}.
Gravitation and Cosmology {\bf 24}, 28 (2018).


\bibitem{my16a} S.A. Pavluchenko,
%{\it Cosmological dynamics of spatially flat Einstein-Gauss-Bonnet models in various dimensions. Vacuum case},
Phys. Rev. D {\bf 94}, 024046 (2016).
%arXiv:1605.01456

\bibitem{my18a} S.A. Pavluchenko, Particles {\bf 1}, 4 (2018)  [arXiv:1803.01887].

\bibitem{my16b} S.A. Pavluchenko,
%{\it Cosmological dynamics of spatially flat Einstein-Gauss-Bonnet models in various dimensions. Vacuum case},
Phys. Rev. D {\bf 94}, 084019 (2016)
%arXiv:1607.07347

\bibitem{my17a} S.A. Pavluchenko, Eur. Phys. J. C {\bf 77}, 503 (2017).

\bibitem{PT2017} S.A. Pavluchenko and A.V. Toporensky, arXiv:1709.04258.

\bibitem{infl1} S.A. Pavluchenko, Phys. Rev. D {\bf 67}, 103518 (2003).

\bibitem{infl2} S.A. Pavluchenko, Phys. Rev. D {\bf 69}, 021301 (2004).

\bibitem{kasner} E. Kasner, Am. J. Math. {\bf 43}, 217 (1921).

\bibitem{Taub} A.H. Taub, Ann. Math. {\bf 53}, 472 (1951).





\bibitem{Tipler} F.J. Tipler, Phys. Lett. {\bf A64}, 8 (1977).

\bibitem{KW1} T. Kitaura and J.T. Wheeler, Nucl. Phys. {\bf B355}, 250 (1991).

\bibitem{KW2} T. Kitaura and J.T. Wheeler, Phys. Rev. D {\bf 48}, 667 (1993).

\end{thebibliography}
\end{document}